\documentclass[aps,prx,reprint,superscriptaddress]{revtex4-1}
\usepackage{lmodern}
\usepackage{graphicx}
\usepackage{braket}
\usepackage{amsfonts, amsmath, amsthm, amssymb}
\usepackage{subfigure}
\usepackage{hyperref}
\hypersetup{colorlinks=true,linkcolor=blue,citecolor=blue,urlcolor=blue}
\def\equationautorefname~#1\null{(#1)\null}
\usepackage{cleveref}
\usepackage[utf8]{inputenc}
\usepackage[T1]{fontenc}
\Crefname{figure}{Fig.}{Figs.}
\newcommand*{\eqnref}[1]{%
\begingroup
Eq. \autoref{#1}%
\endgroup}
\begin{document}

\title{Optical Lattice with Torus Topology}

\author{Hwanmun Kim}
\affiliation{Joint Quantum Institute, NIST/University of Maryland, College Park, Maryland 20742, USA}
\affiliation{Department of Physics, University of Maryland, College Park, Maryland 20742, USA}
\author{Guanyu Zhu}
\affiliation{Joint Quantum Institute, NIST/University of Maryland, College Park, Maryland 20742, USA}
\author{J. V. Porto}
\affiliation{Joint Quantum Institute, NIST/University of Maryland, College Park, Maryland 20742, USA}
\author{Mohammad Hafezi}
\affiliation{Joint Quantum Institute, NIST/University of Maryland, College Park, Maryland 20742, USA}
\affiliation{Department of Physics, University of Maryland, College Park, Maryland 20742, USA}
\affiliation{Departments of Electrical and Computer Engineering and Institute for Research in Electronics and Applied Physics, University of Maryland, College Park, Maryland 20742, USA}

\date{\today}

\begin{abstract}
We propose an experimental scheme to construct an optical lattice where the atoms are confined to the surface of a torus. This construction can be realized with spatially shaped laser beams which could be realized with recently developed high resolution imaging techniques. We numerically study the feasibility of this proposal by calculating the tunneling strengths for atoms in the torus lattice. To illustrate the non-trivial role of topology in atomic dynamics on the torus, we study the quantized superfluid currents and fractional quantum Hall (FQH) states on such a structure. For FQH states, we numerically investigate the robustness of the topological degeneracy and propose an experimental way to detect such a degeneracy. Our scheme for torus construction can be generalized to surfaces with higher genus for exploration of richer topological physics.
\end{abstract}

\pacs{}
\maketitle
\textit{Introduction.}\textbf{\textemdash}
In the past decades, ultracold atoms in optical lattices have been widely used to study a range of interesting coherent and many-body physics \cite{jaksch1998cold}. In particular, there has been remarkable progress in investigating phenomena \cite{greiner2002quantum,aidelsburger2013realization,miyake2013realizing,paredes2004tonks} in both different dimensions \cite{greiner2001bose,stoferle2004transition,spielman2007mott} and lattice geometries, such as square \cite{greiner2001bose,greiner2002quantum}, triangular \cite{becker2010ultracold}, honeycomb \cite{tarruell2012creating}, kagome \cite{jo2012ultracold}, ring \cite{ramanathan2011superflow}, cylinder \cite{lkacki2016quantum}, and more recently ribbon lattices with synthetic dimensions \cite{stuhl2015visualizing}.

Meanwhile, intriguing physics can be explored in systems with non-trivial topologies. For example, it is theoretically predicted that there are topologically protected degeneracies on surfaces with non-zero genus, like the fractional quantum Hall (FQH) model \cite{haldane1985many,wen1990ground} or spin liquids \cite{kitaev2003fault,wen2002quantum,kalmeyer1989theory}. Such systems are expected to not only contain rich many-body physics but also possibly be used in topological quantum computation \cite{kitaev2003fault}. While there have been interesting proposals to make torus surfaces in ultracold atomic systems, using synthetic dimensions \cite{boada2015quantum} and semi-2D geometries by modifying cylinders \cite{grusdt2014realization,budich2017coupled}, the experimental construction of a torus in real space has remained challenging. Moreover, the presence of edge physics and the finite size effect have made the observation of FQH effect in ultracold atoms challenging.

\begin{figure}[t]
\centering
\includegraphics[width=\linewidth]{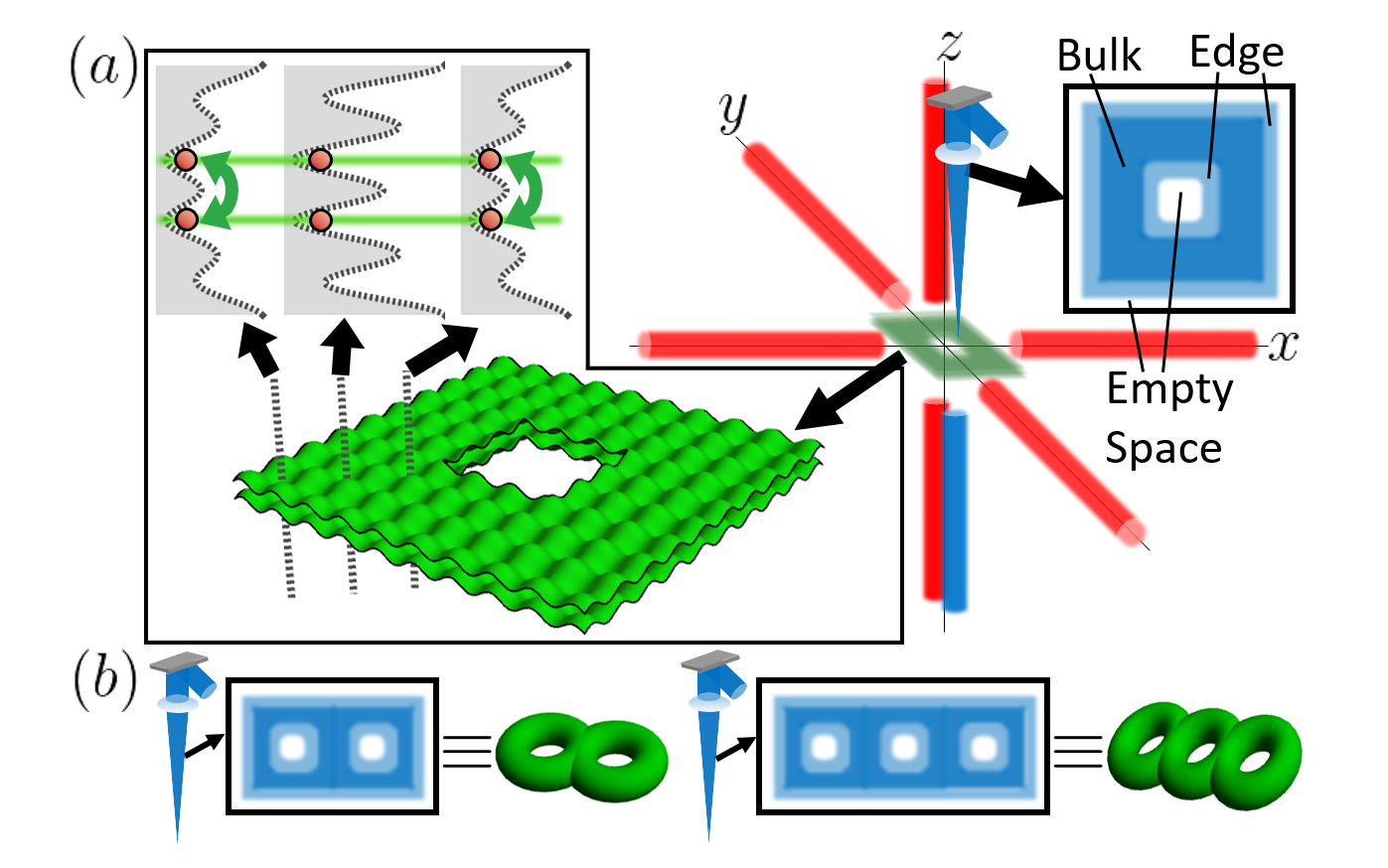}\subfigure{\label{beam_design}}\subfigure{\label{highgenus}}
\caption{(a) Schematic beam configuration for a torus surface in an optical lattice. Plane wave beams in the horizontal directions generate a rectangular lattice in the $xy$ plane. In the $z$ direction, a superlattice structure created by pairs of blue-detuned and red-detuned beams confines atoms in two layers. The $-z$ propagating blue-detuned beam has the beam shape of a square annulus. (Inset) Different laser intensities turn the inter-layer tunneling on and off in different regions. To complete the torus surface, only the inter-layer tunneling on in the edge region is allowed.
(b) Generalization of the scheme to surfaces with higher genus ($g=2,3$ shown for example) can be achieved by puncturing more holes in the middle of the lattice.}\label{beam_scheme}
\end{figure}

In this Letter, we propose a scheme to construct an optical lattice in which atomic dynamics is confined to the surface of a torus. Our construction makes use of recent advances in beam shaping, in the context of ultracold atomic systems \cite{zupancic2016ultra,barredo2016atom,endres2016atom,schine2018measuring,barredo2017synthetic}. Specifically, we show that a rectangular square lattice with a hole in the middle can be turned into the surface of a torus by shaping a single beam perpendicular to the layers (\Cref{beam_scheme}). Moreover, we discuss that this construction could be generalized to surfaces with higher genus. To illustrate the non-trivial role of topology in atomic dynamics on the torus, we first investigate the hydrodynamics of bosonic superfluid on the torus. Specifically, we demonstrate a sequence of optical manipulations that generates quantized supercurrents in two intersecting non-contractible cycles. Furthermore, in the strongly correlated regime, we discuss a FQH model which can be realized on this torus. To numerically investigate the topological degeneracy on such system, we consider a relatively small square lattice ($6 \times 6$) with torus topology. We show that the anticipated topological degeneracy exists and is robust against the discrepancy between inter- and intra-layer tunneling and disorder. Moreover, we  propose a way to experimentally detect the topological degeneracy.

\textit{Torus Construction.}\textbf{\textemdash}
In the following, we show that by using several pairs of laser beams in the $x$, $y$, and $z$ directions, one can build an optical lattice in which atomic dynamics is confined to the surface of a torus (\Cref{beam_scheme}). We first make a bilayer system by creating a superlattice structure in the $z$ direction. Using high resolution optics, we then tailor one of the beams used in the superlattice structure to have the shape of a square annulus. This square annulus divides the $xy$ plane into three regions: \textit{bulk}, \textit{edge}, and \textit{empty space} [\Cref{beam_design}]. By having a different set of intensities in these regions, the trap potential can be arranged to only allow atoms to vertically tunnel through lattice sites in the edge region, thus confining atoms to the surface of a torus.

\begin{figure}[t]
\centering
\includegraphics[width=\linewidth]{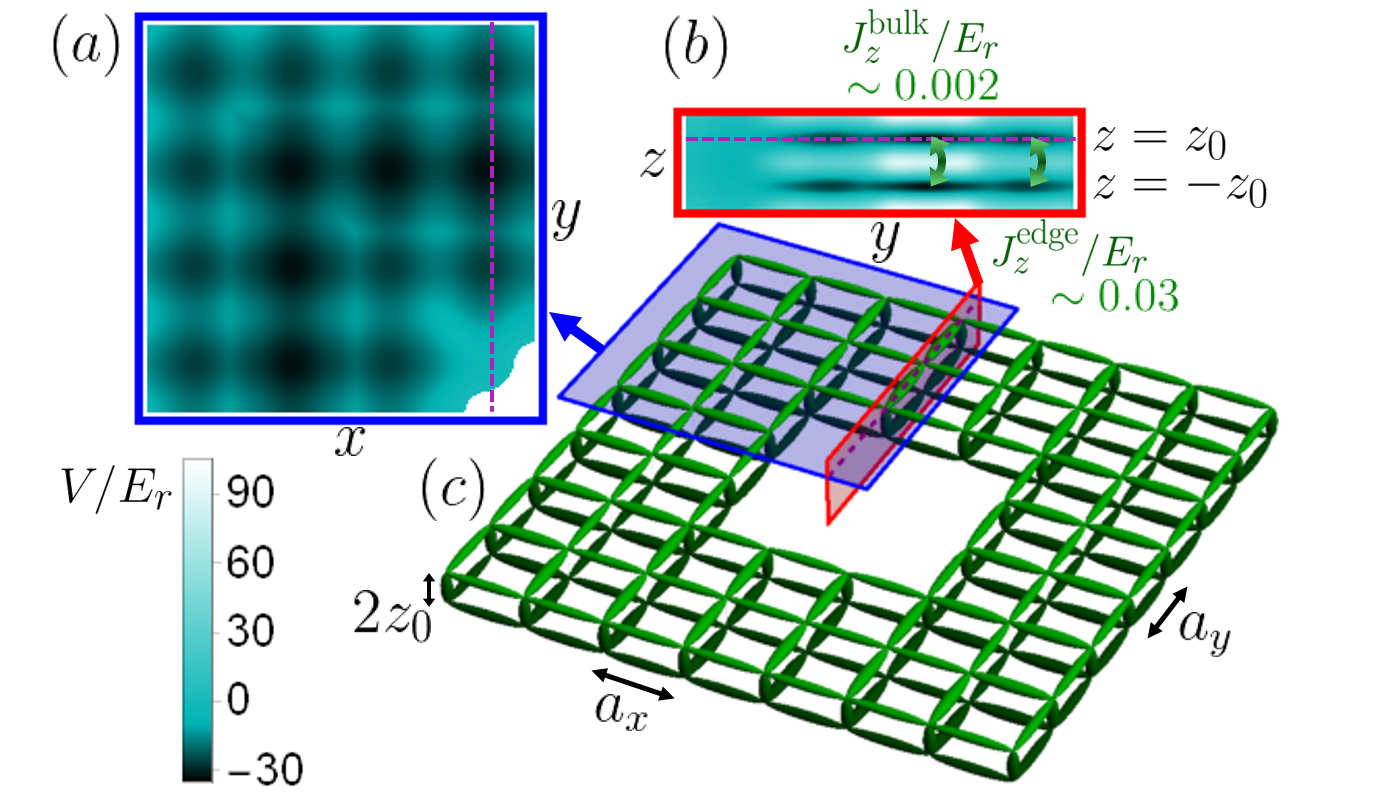}\subfigure{\label{numerical_hopping_a}}\subfigure{\label{numerical_hopping_b}}\subfigure{\label{numerical_hopping_c}}
\caption{Numerically evaluated dipole potential and tunneling strengths. We consider $\text{Rb}^{87}$ atoms with $a_x \simeq a_y=\text{480 nm}$ and $k_x = k_z/2 = 2q_z$. In the unit of recoil energy $E_r \equiv \hbar^2 k_x^2/2m$ ($E_{r,z} \equiv \hbar^2 k_z^2/2m$), $V_{0} = 8 E_r, V_{\text{E}} = 60 E_r (15 E_{r,z}), V_{\text{B}} = 120 E_r (30 E_{r,z}),$ and $V_{\text{red}}=20 E_r (5 E_{r,z})$.
(a) Dipole potentials in the $xy$ plane on the upper layer.
(b) Dipole potentials in the $yz$ plane. Inter-layer tunneling strengths in bulk ($J^{\text{bulk}}_{z}$) and edge ($J^{\text{edge}}_{z}$) are shown for comparison.
(c) Numerically evaluated tunneling strengths represented as the thickness of bonds in the 3D lattice. Shown tunneling strengths range from $0.03 E_r$ to $0.04 E_r$.}\label{numerical_hopping}
\end{figure}

To prepare a bilayer system, we use a 3D optical lattice with a superlattice structure in the $z$ direction. Red-detuned laser beams with wavevectors $\pm k_x \mathbf{\hat{x}}$ and $\pm k_y \mathbf{\hat{y}}$ form a 2D rectangular lattice with lattice spacings $(a_x,a_y)=(\pi/k_x,\pi/k_y)$. For the superlattice structure, we use a pair of blue-detuned lasers with wavevectors $\pm k_z \mathbf{\hat{z}}$ and another pair of red-detuned lasers with wavevectors $\pm q_z \mathbf{\hat{z}}$. When the $\pm z$ propagating beams do not vary in the $xy$ plane, the combined vertical dipole potential is given by $V_z(z)=V_{b}(z)+V_{r}(z)=V_{\text{blue}} \cos^2(k_z z) - V_{\text{red}} \cos^2(q_z z)$ for properly chosen relative phases, where $V_{\text{blue}}$ ($V_{\text{red}}$) is the amplitude of the dipole potential generated by the blue-detuned (red-detuned) beam pair alone. Then atoms with atomic mass $m$ can be confined at two neighboring minima, which we call the $\pm z_0$, as shown in \Cref{numerical_hopping_b}. Atoms in these minima constitute the bilayer system.

In order to complete the torus surface, we tailor the $-z$ propagating blue-detuned beam in the shape of a square annulus in the $xy$ plane, adjusted to achieve the desired inter-layer tunneling only along edge sites. In particular, we make the laser intensity lower at the edge compared to the bulk region. The resulting potential barrier in the $z$ direction is shallower at the edge than the bulk, which makes the inter-layer tunneling non-zero at the edge while negligible in the bulk region. With the laser intensity of the $-z$ propagating beam set to zero in the empty space region, the $+z$ propagating blue-detuned beam generates a higher dipole potential in the empty space compared to the edge and the bulk region. This difference in dipole potential energetically prevents atoms from escaping the designated square annulus.

To be concrete, we consider the following beam shapes for the blue-detuned beams:
\begin{eqnarray}\label{modifiedBeam}
&& \mathbf{E}_+(\mathbf{r},t) = \mathbf{\hat{y}}\left(e^{+ik_z(z-ct)} + \text{c.c.} \right) \mathcal{E}_+, \\
&& \mathbf{E}_-(\mathbf{r},t) =  \mathbf{\hat{y}}\left(e^{-ik_z(z+ct)} + \text{c.c.} \right) 
\left\lbrace \begin{array}{cc}
\mathcal{E}_\text{B} & \text{bulk}\\
\mathcal{E}_\text{E} & \text{edge}\\
0 & \text{empty space}
\end{array} \right. . \nonumber
\end{eqnarray}
In this discrete setting, bulk and edge regions correspond to the zones around bulk and edge sites in the square annulus, within the distance $a_x/2$ ($a_y/2$) in the $x$ ($y$) direction. The rest of the area is designated as empty space. For illustrative purposes, we assume the model beam has sharp boundaries between different regions, but in an experimental realization, one can relax this constraint and construct a good approximation of \eqnref{modifiedBeam} using beams with sufficient numerical apertures (0.17 to 0.80) \cite{supplement}. The recent progress in beam-shaping techniques for optical lattices \cite{zupancic2016ultra,barredo2016atom,endres2016atom,schine2018measuring,barredo2017synthetic,mreed} could allow one to realize such a beam profile in the lab. Note that this beam profile should be placed properly in the $xy$ plane in a way that regional distinctions in \eqnref{modifiedBeam} to match with the horizontal lattice sites.

This beam profile gives rise to the combined vertical dipole potential including interference between the $+z$ and $-z$ propagating beams:
\begin{eqnarray}\label{Vz}
V_z(\mathbf{r}) &=& V_b(\mathbf{r}) - V_{\text{red}} \cos^2(q_z z), \\
V_{b}(\mathbf{r}) &=& \left\lbrace \begin{array}{cc}
V_\text{B}\cos^2(k_z z) + V_\text{B}^{(0)} & \text{bulk}\\
V_\text{E}\cos^2(k_z z) +V_\text{E}^{(0)}  & \text{edge}\\
V_\text{S}^{(0)} & \text{empty space}
\end{array} \right. , \nonumber
\end{eqnarray}
where the lattice potential amplitudes are $V_\text{B/E} \propto 4\mathcal{E}_+ \mathcal{E}_\text{B/E}$, and the energy offsets are $V^{(0)}_{\text{B/E}}\propto(\mathcal{E}_+ - \mathcal{E}_\text{B/E})^2$, $V_\text{S}^{(0)} \propto\mathcal{E}_+^2$. The proportionality constant depends on beam frequency, dipole elements, and transition frequency \cite{grimm2000optical}. By setting $\mathcal{E}_\text{B}>\mathcal{E}_\text{E}$, the potential barrier between layers in the edge region is shallower than in the bulk region. This barrier difference leads to an inter-layer tunneling strength which is stronger in the edge than in the bulk. Moreover, we need to satisfy two additional conditions: (1) to have a smooth torus, the on-site energy in the edge and the bulk regions should be the same, and (2) this on-site energy should be smaller than the potential in the empty space, so that atoms are trapped in the designated square annulus. To find on-site energies in these conditions, we should include the zero point energies in the effective potentials as well. Then, these requirements can be summarized as  
\begin{eqnarray}\label{onsiteconditions}
&&V_\text{B}^{(0)}+ \frac{\hbar\omega_{B}}{2} = V_\text{E}^{(0)} + \frac{\hbar\omega_{\text{E}}}{2} < V_\text{S}^{(0)} 
\end{eqnarray}
where the zero point energy of the harmonic confinements are $\frac{\hbar}{2} \omega_\text{B/E} = \frac{\hbar}{2}\sum_{s=x,y,z} \sqrt{ m^{-1} \left.\partial_s^2 V(\mathbf{r})\right|_{\mathbf{r} \in \text{B/E}} }$. To evaluate this, we consider the total dipole potential $V(\mathbf{r}) = V_{xy}(x,y)+V_z(\mathbf{r})$, where the horizontal dipole potential is $V_{xy}(x,y)=V_{0}\lbrace\cos^2(k_x x)+\cos^2(k_y y)\rbrace$. While it is not obvious to find a set of parameters satisfying these conditions simultaneously, it is possible to satisfy \eqnref{onsiteconditions} by tuning $V_0, V_E, V_B,$ and $V_\text{red}$ only \cite{supplement}.

To verify that our beam design leads to the desired optical lattice, we numerically evaluate the total dipole potential for $\text{Rb}^{87}$ atoms [\Cref{numerical_hopping_a,numerical_hopping_b}]. We approximately evaluate the tunneling strengths by solving the Schr{\"o}dinger equation over the region containing each pair of the nearest neighboring sites \cite{supplement}. \Cref{numerical_hopping_c} shows that it is possible to suppress inter-layer tunneling in the bulk, while simultaneously setting inter-layer tunneling in the edge and intra-layer tunneling everywhere to be non-vanishing. Here, for boundaries between the different regions, we use more realistic resolution limited potentials \cite{supplement} instead of the step functions in \eqnref{modifiedBeam}.

Once our scheme for torus construction is realized, it is straightforward to extend the scheme to surfaces with higher genus [\Cref{highgenus}]. The only requirement is to puncture more holes in the middle of the lattice, which requires no higher resolution in beam-shaping than puncturing a single hole. By generating surfaces with higher genus, one can explore richer topological physics, as we discuss later.

\begin{figure}
\centering
\includegraphics[width=\linewidth]{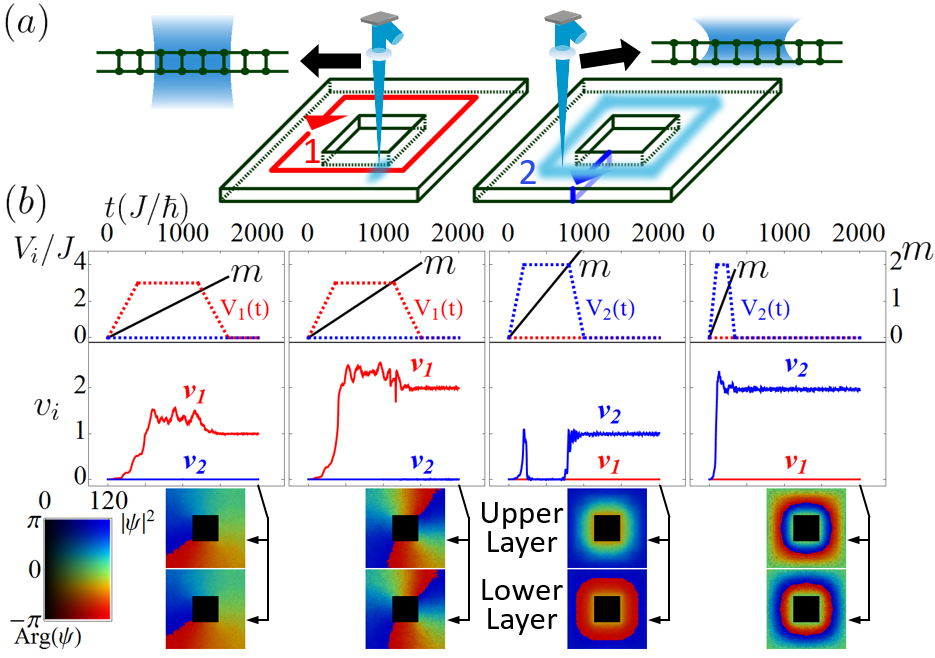}\subfigure{\label{stirring_numeric_a}}\subfigure{\label{stirring_numeric_b}}
\caption{(a) A scheme to generate supercurrents in two cycles. A focused, blue-detuned laser beam acts as a stirrer along each cycle, namely, cycle 1 and 2. Note that the stirrer along cycle 2 is focused on the upper layer. A uniform condensate is loaded on the torus initially, then the stirring potential along cycle 1 ($V_1$) or cycle 2 ($V_2$) is ramped up and down.
(b) Quantization of vorticity in two cycles. Dotted curves in the upper plots indicate the ramping sequences of $V_1$ and $V_2$. Solid lines in the upper plots indicate the number of completed cycles ($m$) in the stirring process. The lower plots show vorticities ($v_i$) changing over time. Steady-state wavefunctions of the different sequences are shown below.}\label{stirring_numeric}
\end{figure}
% ($\uparrow/\downarrow$ for the upper/lower layer)
\textit{Quantized Supercurrents in Two Cycles.}\textbf{\textemdash} 
To demonstrate how topology plays a non-trivial role in the dynamics of ultracold atoms on a torus surface, we numerically investigated the hydrodynamics of weakly interacting bosonic superfluids. Previously, in a ring geometry, it has been experimentally demonstrated that the flow of supercurrents is quantized along the single quantization axis \cite{ryu2007observation,ramanathan2011superflow}. The quantization of supercurrent results from the fact that wavefunction of the atomic condensate should be single-valued and its phase should be compact on a closed cycle. More interestingly, in the torus setting, there are two intersecting non-contractible cycles [\Cref{stirring_numeric_a}] which allow supercurrents to be quantized separately along each. In particular, the vorticity, which is defined as
\begin{eqnarray}\label{vorticity}
v_i = \frac{1}{2\pi \rho_{\text{avg}}} \oint_{\text{cycle }i} \text{Im}\left( \psi^* \mathbf{\nabla} \psi \right)\cdot\mathbf{dl} \quad (i=1,2),
\end{eqnarray}
is quantized to an integer, up to a small finite-size fluctuation. Here, $\rho_{\text{avg}}$ is the average condensate density and $\psi(\mathbf{r})$ is the condensate wavefunction. To generate the supercurrents with non-zero vorticities, we stir the atomic condensate with an extra dipole potential \cite{wright2013threshold}. In particular, we prepare a blue-detuned, focused beam and move it along each non-contractible cycle to generate the supercurrent flow in the stirring direction [\Cref{stirring_numeric_a}]. The supercurrent flows can be detected through established methods, such as time-of-flight imaging \cite{ryu2007observation}.

To specifically show the quantization along the two non-contractible loops, we numerically simulate these stirring procedures [\Cref{stirring_numeric_b}]. In the weakly interacting and tight-binding regime, atomic dynamics in our optical lattice can be described in the mean-field approximation, 
\begin{eqnarray}\label{gpeeqn}
i\hbar \partial_t \psi^{\uparrow/\downarrow}_{j} &=& -J \sum_{k; |k-j|=1} \psi^{\uparrow/\downarrow}_{k} - \left(J \psi^{\downarrow/\uparrow}_{j}  \right) \delta_{j\in\text{edge}} \nonumber\\
&& + \left\lbrace V^{\uparrow/\downarrow}(\mathbf{r}_j,t)-\mu +U \left|\psi^{\uparrow/\downarrow}_{j} \right|^2 \right\rbrace \psi^{\uparrow/\downarrow}_{j},
\end{eqnarray}
where $\psi^{l}_{j}$ is the condensate wavefunction at site $j$ on layer $l$, which can be $\uparrow/\downarrow$ for the upper/lower layer. In this equation, $|k-j|$ indicates the distance between site $k$ and $j$ within the same layer, while $\delta_{j\in\text{edge}}$ is $1$ if $j$ belongs to the edge region and $0$ otherwise. $J$ is the tunneling strength, $U$ is the on-site interaction energy, $V^{l}$ is the stirring potential on layer $l$ and $\mu$ is the chemical potential. This dynamics can be simulated with the numerical methods for the Gross-Pitaevskii equation \cite{gross1961structure,pitaevskii1961vortex,bao2003numerical}. See \cite{supplement} for further details of the simulation.

In the simulation, we verify that the stirred superfluid exhibits the quantized vorticity along each cycle of stirring [\Cref{stirring_numeric_b}]. We also see that this vorticity increases with the stirring speed. As expected, the evaluated vorticity along each cycle coincides with the wavefunction winding numbers [\Cref{stirring_numeric_b}]. Also, we observe the creation and annihilation of vortex-antivortex pairs during the increment of vorticity \cite{supplement}.

\begin{figure}[t]
\centering
\includegraphics[width=\linewidth]{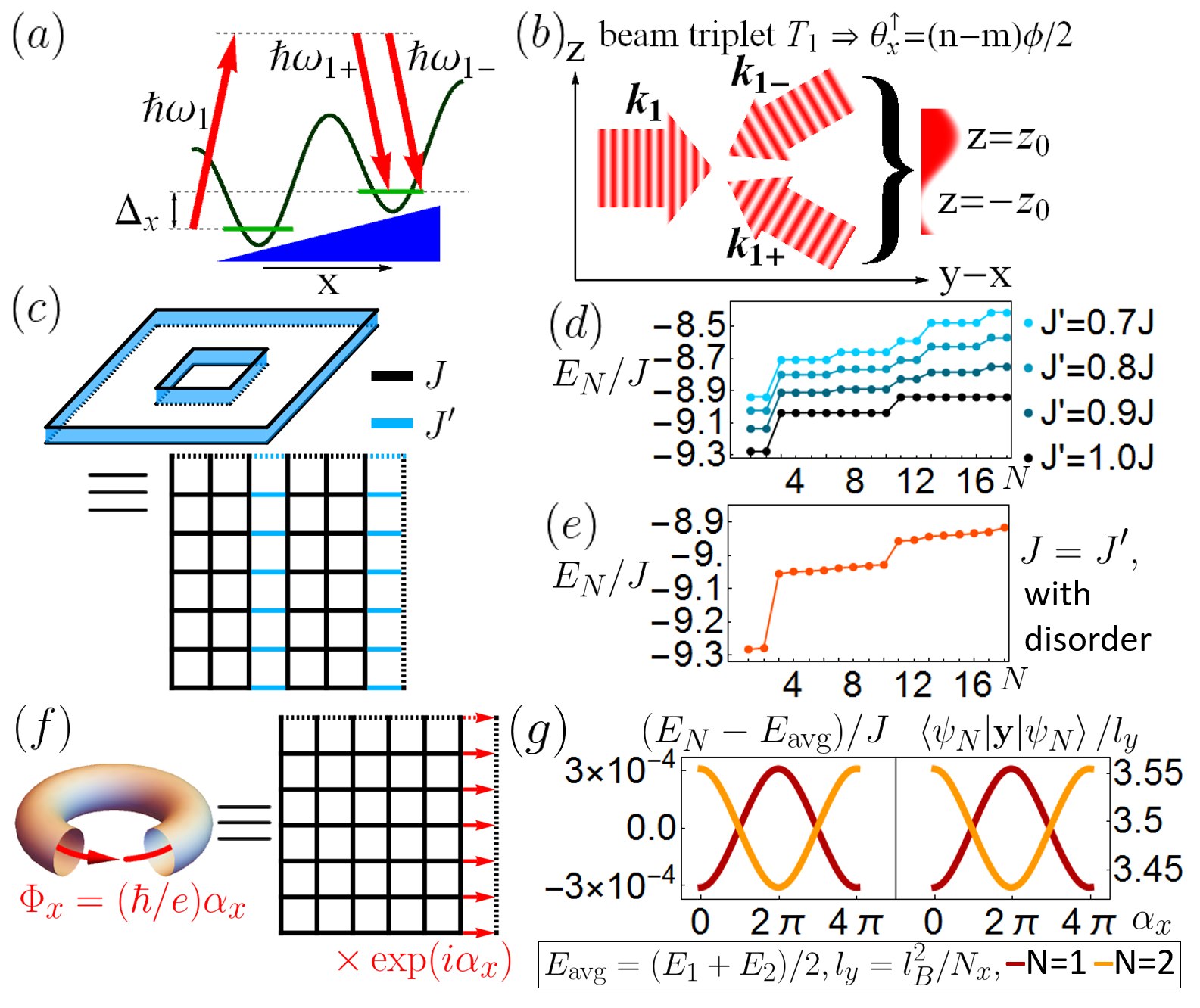}\subfigure{\label{fqhe_a}}\subfigure{\label{fqhe_b}}\subfigure{\label{fqhe_c}}\subfigure{\label{fqhe_d}}\subfigure{\label{fqhe_e}}\subfigure{\label{fqhe_f}}\subfigure{\label{fqhe_g}}
\caption{(a),(b) A scheme for FQH Hamiltonian. Different Raman beam triplets $T_{\mathbf{1}} \sim T_{\mathbf{4}}$ give the different tunneling phases in \eqnref{fqh_hamiltonian}. Schematic beam configuration of $T_{\mathbf{1}}$ is shown for an example. Zeeman eneargy difference $\Delta_x$ ($\Delta_y$) in the $x$ ($y$) direction is matched with detuning of Raman beams in triplets $T_{\mathbf{1}}$ and $T_{\mathbf{3}}$ ($T_{\mathbf{2}}$ and $T_{\mathbf{4}}$) to give tunneling terms in the same direction. To address each layer independently, beam $\mathbf{i+}$ and $\mathbf{i-}$ among triplet $T_{\mathbf{1}}$ or $T_{\mathbf{2}}$ ($T_{\mathbf{3}}$ or $T_{\mathbf{4}}$) destructively interfere at lower (upper) layer. (c) Exact diagonalization of FQH Hamiltonian for 3 hardcore bosonic atoms on a $6\times6$ square lattice ($N_x=N_y=6$) with periodic boundary conditions and $\phi=\pi/3$, magnetic length $l_B \equiv \sqrt{2\pi/\phi}$. (d) Energy spectrums with distinct intra-layer $(J)$ and inter-layer $(J')$ tunnelings. (e) Spectrum with a random disorder of scale $0.05J$. Energy splitting between the ground states is $5\times 10^{-3} J$. (f) Inserting flux $\Phi_x$ through the handle of torus is equivalent to the boundary condition with twist angle $\alpha_x$. (g) With additional potential $V(y) = (0.01J/N_y)y$, the spectral flows in $\alpha_x$ can be detected by measuring the $y$-coordinates of the states.
}\label{fqhe}
\end{figure}

\textit{Topological Degeneracy in FQH States.}\textbf{\textemdash} 
Our construction allows one to investigate the dynamics of strongly interacting ultracold atoms on a torus. As an example, we study a bosonic FQH model, which could be realized by laser-assisted tunneling \cite{aidelsburger2013realization,miyake2013realizing}. Specifically, the lattice FQH Hamiltonian for bosonic atoms on our torus can be written as 
\begin{eqnarray}\label{fqh_hamiltonian}
&& H = \sum_{n,m}\sum_{l=\uparrow,\downarrow} \left(\frac{U}{2} {a^{l\dag}_{n,m}}^2 {a^{l}_{n,m}}^2  \right. \nonumber\\
&&\qquad\qquad \left. -J e^{i\theta^{l}_{x}} a^{l\dag}_{n+1,m} a^{l}_{n,m} -J e^{i\theta^{l}_{y}} a^{l\dag}_{n,m+1} a^{l}_{n,m} +\text{H.c.} \right) \nonumber\\
&&\qquad\qquad -\sum_{(n,m)\in \text{edge}} \left( J' a^{\uparrow\dag}_{n,m} a^{\downarrow}_{n,m} + \text{ H.c.} \right) , \\
&&\text{where } \theta^{\uparrow/\downarrow}_{x}(n,m) = \frac{(n \mp m)\phi}{2}, \ \theta^{\uparrow/\downarrow}_{y}(n,m) = \frac{(m \pm n)\phi}{2} \nonumber.
\end{eqnarray}
Here, $a^{l}_{n,m}$ annihilates an atom at site $(n,m)$ on layer $l$. $J$ and $J'$ are the effective intra- and inter-layer tunneling strengths, and $U$ is the on-site interaction energy. With proper size of square annulus, the synthetic magnetic flux per unit cell can be set to $\phi$ \cite{supplement}. To obtain the tunneling phases in \eqnref{fqh_hamiltonian}, we apply a magnetic field in such a way that the Zeeman energy gradient becomes $\Delta_x$ ($\Delta_y$) per site in the $x$ ($y$) direction. Then we apply Raman beams whose detuning matches with $\Delta_x$ ($\Delta_y$) to induce the tunneling in the $x$ ($y$) direction [\Cref{fqhe_a}]. Since the surface orientations of two layers are opposite to each other, the required tunneling phases in each layer should be different as well. This can be achieved by targeting the different Raman beams on the different layers [\Cref{fqhe_b}]. To do so, we use a triplet of beams for each tunneling term, namely $T_{\mathbf{i}} \equiv \lbrace\mathbf{i},\mathbf{i+},\mathbf{i-}\rbrace$, $i=1$ to $4$. Here, the beam $\mathbf{i}$ ($\mathbf{i \pm}$) has the frequency $\omega_i$ ($\omega_{i\pm}$) and the wavevector $\mathbf{k_{i}}$ ($\mathbf{k_{i\pm}}$). In this triplet, the beams $\mathbf{i+}$ and $\mathbf{i-}$ have the same $x$ and $y$ components and have the opposite $z$ components in the wavevectors. These two beams then form a standing wave in $z$ direction. By aligning the beams $\mathbf{i+}$ and $\mathbf{i-}$ to destructively interfere at the lower (upper) layer, the beam triplet $T_{\mathbf{i}}$ can solely address the upper(lower) layer. In a rotating frame, these Raman beams result in the effective tunneling terms given in \eqnref{fqh_hamiltonian} \cite{supplement}.

We numerically investigated the topological degeneracy in FQH system on the torus. In particular, FQH systems with filling fraction $\nu = 1/m$ on a torus surface have $m$-fold ground state degeneracies \cite{haldane1985many,wen1990ground}. To numerically diagonalize the FQH Hamiltonian, we put the upper layer part of Hamiltonian in \eqnref{fqh_hamiltonian} on a $6\times 6$ square lattice with periodic boundary conditions [\Cref{fqhe_c}]. For the filling fraction $\nu=1/2$, we have the anticipated two-fold degeneracy in the ground states [\Cref{fqhe_d}].

To examine the robustness of this degeneracy, we diagonalize the same Hamiltonian with distinct intra-layer ($J$) and inter-layer ($J'$) tunnelings. As seen in \Cref{fqhe_d}, the two-fold degeneracy persists while $J'$ varies from $J$ to $0.7J$. As another test for the robustness of this degeneracy, we also diagonalize the same system with a disorder potential [\Cref{fqhe_e}]. Here, we observe a slight splitting between the ground states, which is much smaller than disorder scale $0.05J$ and the excitation gap. Therefore, this topological degeneracy in a small FQH system is robust against potential experimental imperfections.

Furthermore, one can measure the topological degeneracy by measuring the spectral flow during the synthetic magnetic flux insertion though the handle of the torus. As shown in \Cref{fqhe_f}, the insertion of flux $\Phi_x$ is equivalent to the boundary condition $\psi(x+N_x,y)=\psi(x,y)\exp(i\alpha_x)$ where $\alpha_x = (e/\hbar)\Phi_x$. For $\nu=1/m$, the spectral flow of each ground state shows the $2m\pi$-periodicity in $\alpha_x$ \cite{hatsugai1991anyons,hafezi2007characterization}. To observe this periodicity, we can introduce a small energy splitting by applying a potential $V(y) \propto y$. Such a spectral flow is manifested in the $y$-coordinate expectation values of the ground states [\Cref{fqhe_g}]\cite{supplement}. This average atom position can be experimentally detected through the density measurements.

\textit{Outlook.}\textbf{\textemdash}
By introducing $g$ punctures, we can generalize our scheme to a genus-$g$ surface and leads to a topologically protected $m^g$-fold degenerate ground-state subspace for abelian and non-abelian FQH states. In that context, one can implement modular transformations to probe topological orders, measure fractional statistics, and realize fault-tolerant logical gates for topological quantum computations \cite{Barkeshli:2016wp, Zhu:2017tr}.

\textit{Acknowledgment.}\textbf{\textemdash}
We would like to thank Peter Zoller, Iacopo Carusotto, Ian Spielman, Maissam Barkeshli, Gretchen Campbell, Michael Foss-Feig, and Hirokazu Miyake for insightful discussions. This research was supported by the Physics Frontier Center at the Joint Quantum Institute.

\begin{appendix}
\begin{widetext}

\section{Validity of Model Laser Beams}

In our design of optical lattice, we assume each laser beam maintains the beam shape in their propagating direction. While such assumption is reasonable for the plane wave beams, the model beam shape \eqnref{modifiedBeam} with this assumption would violate Maxwell's laws. If we modify this model beam to satisfy Maxwell's laws, the beam shape should change as the beam propagates. For this modified beam to be a good approximation of \eqnref{modifiedBeam}, we should check the change of beam shape is modest over the region in which our bilayer system is located.
On the other hand, beam shaping with a high precision requires experimental schemes to focus laser beams in the targeted area. To make sure that our beam design is experimentally feasible, we should check if the highest numerical aperture (NA) required in our design is achievable with the current technology.

To construct an approximation of the $-z$ propagating beam ($\mathbf{E}_{-}$) in \eqnref{modifiedBeam}, in a way that Maxwell's laws are satisfied, we reconstruct the 3D intensity profile of this model beam with Hermite-Gaussian (HG) decomposition (\Cref{HG}). Each HG mode has the form of \cite{siegmanlasers}
\begin{eqnarray}\label{HGexpression}
\mathcal{E}_{lm}(x,y,z) &=& \mathcal{E}_0\frac{w_0}{w(z)} H_l\left(\frac{\sqrt{2}x}{w(z)}\right) H_m\left(\frac{\sqrt{2}y}{w(z)}\right) \exp\left( -\frac{x^2+y^2}{w^2(z)} \right) \exp\left( -i\frac{\pi(x^2+y^2)z}{\lambda(z^2 + (\pi w_0^2/\lambda)^2)} -i\frac{2\pi z}{\lambda}  + i\eta(z)\right), \nonumber\\
&& w(z) = \sqrt{w_0^2 + (\lambda z/\pi w_0)^2}, \ \eta(z) = (l+m+1) \tan^{-1}\left( \frac{z^2}{z^2 + (\pi w_0^2/\lambda)^2} \right).
\end{eqnarray}
Here, $H_l(x)$ is the $l$th order Hermite polynomial, $\lambda$ is the wavelength of the beam and $w_0$ is the beam waist radius. Since each mode is a solution of the electromagnetic wave equation, any superposition of concentric and confocal HG modes satisfy Maxwell's laws. In particular, we consider the superposition such that 
\begin{eqnarray}\label{decomposition}
\left| \mathbf{E}_{-} (\mathbf{r},t) \right|_{z=0,t=0}
= \sum_{l,m=0}^{M} C_{lm} \mathcal{E}_{lm}(x,y,0).
\end{eqnarray}
Since $|\mathbf{E}_{-}|$ is an even function in $x$ and $y$, we can omit modes with odd $l$ or odd $m$. While full HG decomposition requires $M \to \infty$, we set $M=120$ to keep the required NA of the beams experimentally accessible. We also need to replace the step functions in $|\mathbf{E}_{-}|$ with smoother functions. In particular, we use sinusoidal functions in the overlapping region between the different regions. For example, if the cut is located at $x=0$, beam amplitude changes as
\begin{eqnarray}
\mathcal{E}(x) = \left\lbrace \begin{array}{cc}
\mathcal{E}_1 & -a_x/2 < x < -b a_x /2 \\
(\mathcal{E}_1-\mathcal{E}_2)\cos^2\left\lbrace(\pi x)/(2 b a_x)+\pi/4\right\rbrace & |x|< b a_x /2 \\
\mathcal{E}_2 & b a_x /2 < x < a_x/2
\end{array} \right.
\end{eqnarray}
where $b=0.7$ between the edge and the empty space, and $b=0.4$ for the rest of boundaries. To carry out the numerical evaluation in \Cref{HG}, we use $\lambda=\text{480 nm}$ and $w_0=\text{1264 nm}$ over the lattice with $a_x=a_y=\text{480 nm}$.  As shown in \Cref{HG_a}, the intensity profile of the reconstructed beam almost maintains its beam shape over the region of our bilayer system. Therefore, one can construct an approximation of the model beam that satisfies Maxwell's laws.

To see if this reconstructed beam in \Cref{HG_a} is achievable with reasonable NA, we calculate numerical aperture of each HG mode used in the reconstructed beam. Since \eqnref{HGexpression} is separable in $x/w(z)$ and $y/w(z)$, we can define the radius of the mode $l$, $r_l(z)$, in a way that the intensity proportion of the HG beam of mode $l,m$ that passes through the ellipse $\lbrace \mathbf{r} | x^2/r^2_l(z) + y^2/r_m(z)^2 \le 1 \rbrace$ is equal to the certain fidelity. By setting this fidelity to be $0.99$, we get
\begin{eqnarray}
\left(\int_{-\infty}^{\infty} H^2_l\left(\frac{\sqrt{2}x}{w(z)}\right) \exp\left(-\frac{2x^2}{w^2(z)}\right)dx\right)^{-1} \int_{-r_l(z)}^{r_l(z)} H^2_l\left(\frac{\sqrt{2}x}{w(z)}\right) \exp\left(-\frac{2x^2}{w^2(z)}\right)dx = \sqrt{0.99}.
\end{eqnarray}
Since $r_l(z)$ is proportional to $w(z)=\sqrt{w_0^2 + (\lambda z/\pi w_0)^2}$, $r_l(z)\sim \alpha_l z$ for $z \gg w_0^2/\lambda$. Then the numerical aperture for the $l$th mode is given by $\text{NA}_l = \sin(\tan^{-1}\alpha_l) = (1+\alpha_l^{-2})^{-1/2}$, which is evaluated in \Cref{HG_b}. As shown in the figure, the numerical apertures of HG modes used in \eqnref{decomposition} range from $\text{NA}=0.17$ to $\text{NA}=0.80$. Since high-order HG beams are already implemented with NA=0.8 \cite{zupancic2016ultra} and the focused beam with NA=0.92 for addressing of ultracold atoms is experimentally reported \cite{robens2017high}, the reconstructed beam in \Cref{HG} is experimentally promising.

\begin{figure}[h]
\centering
\includegraphics[width=\linewidth]{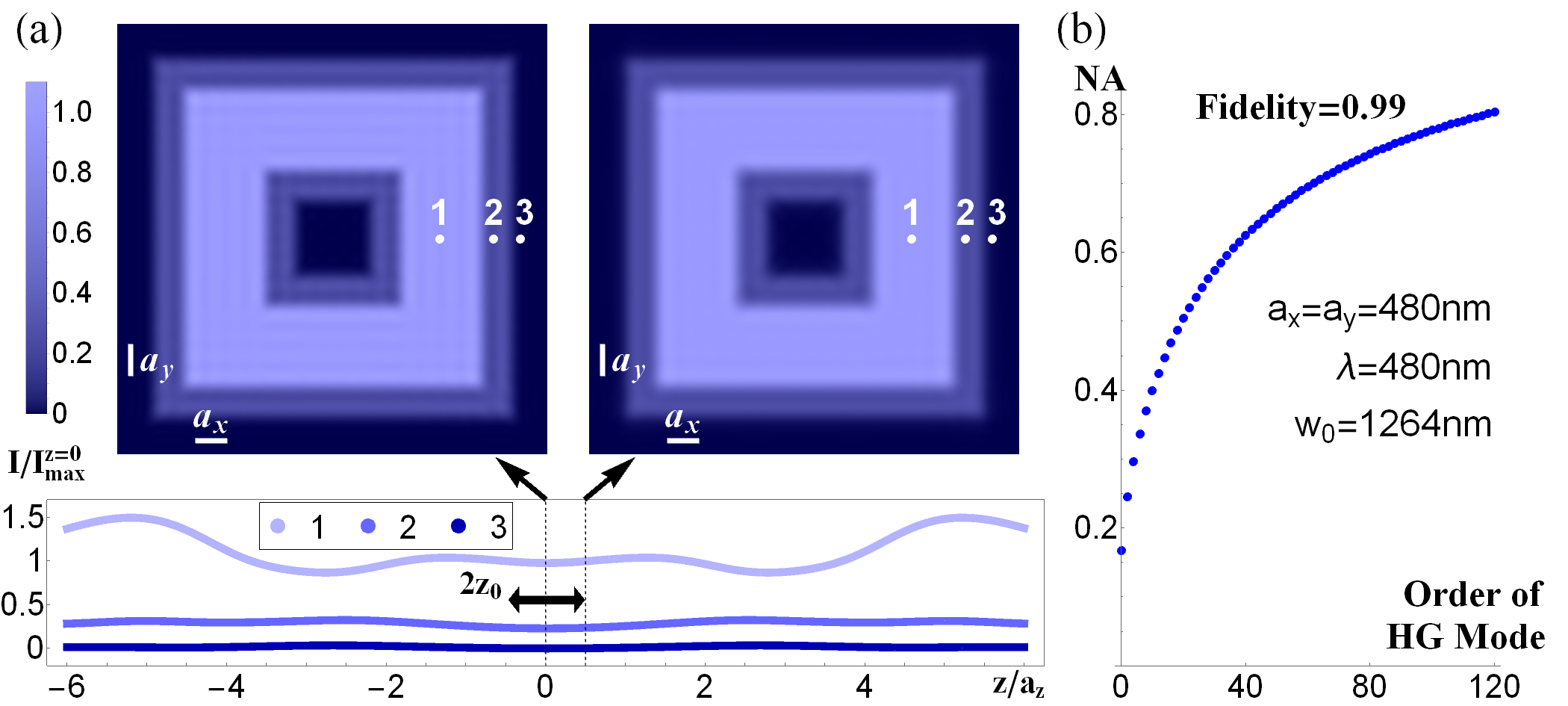}\subfigure{\label{HG_a}}\subfigure{\label{HG_b}}
\caption{(a) Intensity profile of the reconstructed beam through HG decomposition. Here, $I_{\text{max}}^{z=0}$ indicates the maximum intensity at $z=0$. The beam shape is almost maintained over $|z|\le z_0$, where the bilayer system are located. (b) Numerical apertures of HG modes used in the reconstructed bean, which ranges from 0.17 to 0.80. We used HG modes up to the order of 120, while only the even modes are used for the symmetry reason.}\label{HG}
\end{figure}

We also use the focused laser beams used in supercurrent generation procedure (\Cref{stirring_numeric}). Since the focused laser beam with NA=0.92 is reported \cite{robens2017high}, we use this number as the benchmark for the stirring laser beams. With the NA=0.92 and the wavelength $\lambda=\text{480 nm}$, the Gaussian beam has the waist radius $w_0 = \lambda/\pi (\text{NA}^{-2}-1)^{0.5}= \text{65 nm}$, which is far smaller than the lattice spacing in the numerical evaluation in \Cref{numerical_hopping}. This tells that the focused beam used in stirring in the cycle 1 is experimentally feasible. For the stirrer along the cycle 2, we need to obtain enough imbalance in intensities of the focused laser beam reaching upper and lower layers. In our numerical evaluation of the optical lattice, the distance between the upper and lower layers is $z_0 = \text{120 nm}$. If this Gaussian beam is focused at one of the layers, the central laser intensity at the other layer is
\begin{eqnarray}\label{mingam}
I_{z_0} = I_0\left\lbrace 1+ \left(\frac{\lambda z_0}{\pi w_0^2}\right)^2 \right\rbrace^{-1} = (0.051) I_0,
\end{eqnarray}
where $I_0$ is the central laser intensity at the focused layer. This provides a lower bound of the intensity ratio $\gamma=I_{z_0}/I_0$ which we introduce and compare later.

\section{Conditions for On-site energies}
In \eqnref{onsiteconditions}, we stated the required conditions for the on-site energies. To satisfy these conditions in the dipole potential plotted in \Cref{numerical_hopping}, we calculate $\mathcal{E}_+, \mathcal{E}_\text{B},$ and $\mathcal{E}_\text{E}$ for given potential parameters $V_\text{E}, V_\text{B},$ and $V_\text{red}$. For this, we introduce the proportionality constant $f_0$ between the dipole potential and the beam intensity. Then we restate \eqnref{onsiteconditions} as
\begin{eqnarray}\label{onsiteconds}
(\mathcal{E}_+ - \mathcal{E}_\text{B})^2 f_0 + \sqrt{\frac{\hbar^2}{4m} \left.\partial^2_z V_z(\mathbf{r})\right|_{\mathbf{r}\in\text{B}}}
= (\mathcal{E}_+ - \mathcal{E}_\text{E})^2 f_0 + \sqrt{\frac{\hbar^2}{4m} \left.\partial^2_z V_z(\mathbf{r})\right|_{\mathbf{r}\in\text{E}}} < \mathcal{E}_{+}^2 f_0
\end{eqnarray}
Here the quantity $\left[(\hbar^2/4m)\left.\partial^2_z V_z(\mathbf{r})\right|_{\mathbf{r}\in\text{B/E}}\right]^{1/2} \equiv E_{\text{B/E}}^{(0)}$ is solely determined by $V_\text{E}, V_\text{B},$ and $V_\text{red}$. Now the equation part of \eqnref{onsiteconds} becomes
\begin{eqnarray}
&& \mathcal{E}_\text{B}^2 - 2 \mathcal{E}_+ \mathcal{E}_\text{B} + E_\text{B}^{(0)}/f_0
= \mathcal{E}_\text{E}^2 - 2 \mathcal{E}_+ \mathcal{E}_\text{E} + E_\text{E}^{(0)}/f_0
\quad\leftrightarrow\quad\frac{V_\text{B}^2}{16\mathcal{E}_+^2 f_0^2}+\frac{E_\text{B}^{(0)} - V_\text{B}/2}{f_0}
= \frac{V_\text{E}^2}{16\mathcal{E}_+^2 f_0^2} + \frac{E_\text{E}^{(0)} - V_\text{E}/2}{f_0} \nonumber\\
&&\leftrightarrow\quad \mathcal{E}_+ = \sqrt{\frac{V_\text{B}^2 - V_\text{E}^2}{16f_0\lbrace (V_\text{B}-V_\text{E})/2 - (E_\text{B}^{(0)}-E_\text{E}^{(0)}) \rbrace}}, \quad \mathcal{E}_\text{B/E} = \frac{V_\text{B/E}}{4f_0 \mathcal{E}_+} =  \sqrt{\frac{ V_\text{B/E}^2 \lbrace (V_\text{B}-V_\text{E})/2 - (E_\text{B}^{(0)}-E_\text{E}^{(0)}) \rbrace}{(V_\text{B}^2 - V_\text{E}^2)f_0}}.
\end{eqnarray}
By plugging these expressions in the inequality of \eqnref{onsiteconds}, we obtain an inequality between $V_\text{E}, V_\text{B},$ and $V_\text{red}$. It is straightforward to check the potential parameters in \Cref{numerical_hopping} satisfies this inequality. With the same set of parameters, we can evaluate the on-site energy for each site and verify that it is the same for all sites (\Cref{onsiteplot}). As shown in the figure, on-site energies in the bulk and the edge are matched so that atoms can tunnel to each other within the same layer. By doing so, one can make the torus surface smooth.

\begin{figure}
\centering
\includegraphics[width=\linewidth]{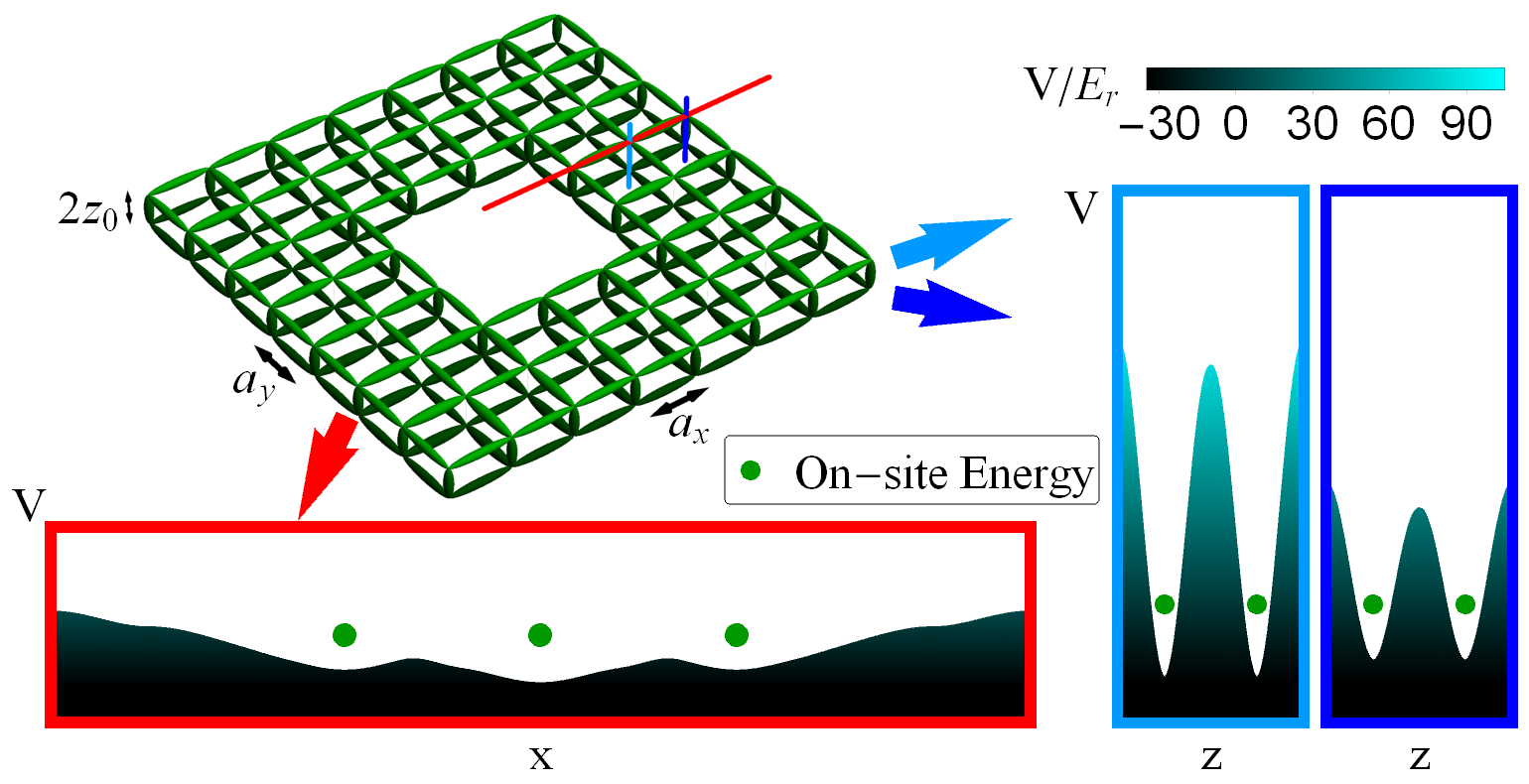}
\caption{
On-site energy presented with the dipole potential along several lines in the optical lattice. On-site energies in the bulk and the edge are set to be equal, which leads to a smooth torus surface.}\label{onsiteplot}
\end{figure}

\section{Numerical Evaluation of Tunneling Strength}
To find the tunneling strength between two neighboring sites, we use the isolated two-site model for this pair of sites. If $E_1$ and $E_2$ are on-site energy site 1 and 2 in this pair, the model Hamiltonian is given by $\left( \begin{array}{cc} E_1 & -J^* \\ -J & E_2 \end{array} \right)$, whose eigenenergies are given by $\epsilon_{\pm} = (E_1+E_2)/2 \pm \sqrt{|J|^2 + (E_1-E_2)^2/4}$. Conversely, the tunneling strength $|J|$ is given by
\begin{eqnarray}\label{tunneling_strength}
|J| = \frac{1}{2}\sqrt{(\epsilon_{+} - \epsilon_{-})^2 - (E_1-E_2)^2}.
\end{eqnarray}

To evaluate $E_1$, $E_2$, and $\epsilon_{\pm}$, we numerically solve the 3D Schrodinger equations. For example, for two sites located at $(x,y,z)=(-a_x/2,0,0)$ and $(x,y,z)=(a_x/2,0,0)$, we take points in the real space as
\begin{eqnarray}
&& x_n = -a_x + (n-1/2)\delta x, \ y_n = -a_y/2 + (n-1/2)\delta y, \ z_n = -a_z/2 + (n-1/2)\delta z, \quad n\in \mathbb{N},\nonumber\\
&& X_1 = \lbrace x_n | n \le N_x \rbrace, \ X_2 = \lbrace x_n | N_x < n \le 2N_x \rbrace, \ Y = \lbrace y_n | n \le N_y \rbrace, \ Z = \lbrace z_n | n \le N_z \rbrace, \nonumber\\
&& \quad\text{where} \ \delta x = a_x/N_x, \ \delta y = a_y/N_y, \ \delta z = a_z/N_z.
\end{eqnarray}
Denote the position $(x,y,z)=(x_i,y_j,z_k)$ as $ijk$. Now the discrete Schrodinger equation gives the following Hamiltonian:
\begin{eqnarray}
H_{i'j'k',ijk} = \left\lbrace \begin{array}{cc}
-(\hbar^2 / 2m)(\delta x)^{-2} & \text{ if } |i'-i|=1, \ j'=j, \ k'=k \\
-(\hbar^2 / 2m)(\delta y)^{-2} & \text{ if } i'=i, \ |j'-j|=1, \ k'=k \\
-(\hbar^2 / 2m)(\delta z)^{-2} & \text{ if } i'=i, \ j'=j, \ |k'-k|=1 \\
(\hbar^2 / m) \left\lbrace (\delta x)^{-2} + (\delta y)^{-2} + (\delta z)^{-2} \right\rbrace + V(x_i,y_j,z_k) & \text{ if } i'=i, \ j'=j, \ k'=k
\end{array}\right. .
\end{eqnarray}
Here, $V(x,y,z)$ is the dipole potential introduced in \eqnref{Vz}. Now $E_1$ ($E_2$) is the lowest eigenenergy obtained by numerically diagonalizing this $H$ over $X_1 \times Y \times Z$ ($X_2 \times Y \times Z$), while $\epsilon_{-}$ and $\epsilon_{+}$ are the first and second lowest eigenenergies obtained by numerically diagonalizing $H$ over $(X_1 \cup X_2) \times Y \times Z$. From \eqnref{tunneling_strength}, we can evaluate the tunneling strength between the two targeted site.

This method can be applied to every pair of neighboring sites, as presented in \Cref{numerical_hopping}. For the tunneling strengths in the figure, we use $N_x = N_y = N_z = 60$ in the evaluation. With the dipole potential presented in \Cref{numerical_hopping}, in the unit of recoil energy $E_r = \hbar^2 k_x^2 /2m$, calculated tunneling strengths are given as follows; the intra-layer tunneling strength between two bulk sites is $0.032 E_r$, the intra-layer tunneling strength between an edge site and a bulk site is $0.041 E_r$, the inter-layer tunneling strength between two edge sites is $0.036 E_r$, and the inter-layer tunneling strength between two bulk sites is $0.002 E_r$.

\section{Numerical Simulation of Dynamics in Condensate With Stirring Potentials}

As presented in \eqnref{gpeeqn}, in the mean field limit, the dynamics of condensate wavefunction is determined by
\begin{eqnarray}
&& i\hbar \partial_t \psi^{\uparrow/\downarrow}_{j} = -J \sum_{k; |k-j|=1} \psi^{\uparrow/\downarrow}_{k} - \left(J \psi^{\downarrow/\uparrow}_{j}  \right) \delta_{j\in\text{edge}} + \left\lbrace V^{\uparrow/\downarrow}(\mathbf{r}_j,t)-\mu+ U \left|\psi^{\uparrow/\downarrow}_{j} \right|^2 \right\rbrace \psi^{\uparrow/\downarrow}_{j},
\end{eqnarray}
where $j$ runs over sites in each layer. Stirring potential $V^{l}(\mathbf{r}_j,t) = V^{l}(x,y,t) = V^{l}(r,\phi,t)$ ($l=\uparrow/\downarrow$ for upper/lower layer) in this equation is given by
\begin{eqnarray}\label{simeqn}
\begin{array}{ccc}
V^{\uparrow}(\mathbf{r}_j,t) &=& V_1(t) e^{-[(x-X(t))^2+(y-Y(t))^2]/2d_1^2} + V_2(t)( e^{-(r-R_A(t))^2/2d_2^2} + \gamma e^{-\gamma(r-R_B(t))^2/2 d_2^2} ) \\
V^{\downarrow}(\mathbf{r}_j,t) &=& V_1(t) e^{-[(x-X(t))^2+(y-Y(t))^2]/2d_1^2} + V_2(t)( \gamma e^{-\gamma(r-R_A(t))^2/2 d_2^2} + e^{-(r-R_B(t))^2/2d_2^2} )
\end{array}.
\end{eqnarray}
Here, $(X(t),Y(t)) = R_0 (\cos (2\pi t/\tau_1),\sin (2\pi t/\tau_1))$ and $R_{A,B}(t) = R_1 + (R_2-R_1)\left(\text{mod}(\pm (t-T_\text{half})/\tau_2,2)\right)$. In each stirring sequence in \Cref{stirring_numeric_b}, $V_i(t)$ ($i=1$ for the first two graphs and $i=2$ for the last two graphs) ramps up from 0 to $V_{\text{max},i}$, then remains at $V_{\text{max},i}$, and finally ramps down to 0.

In this simulation, we consider a torus embedded in two layers of $108\times 108$ square lattice ($a_x=a_y=a$) with a $36\times 36$ puncture in the middle. Numerical parameters used in this simulation are $U=0.0041J, \mu=27.43J, V_{\text{max},1}=3.0J, V_{\text{max},2}=4.0J, R_0=36.0a, R_1=32.0a, R_2=80.0a, d_1=12.0a, d_2=4.0a, \gamma=0.2$. Note that $\gamma$ used in this simulation is well above the lower bound obtained in \eqnref{mingam}.

\begin{figure}[h]
\centering
\includegraphics[width=\linewidth]{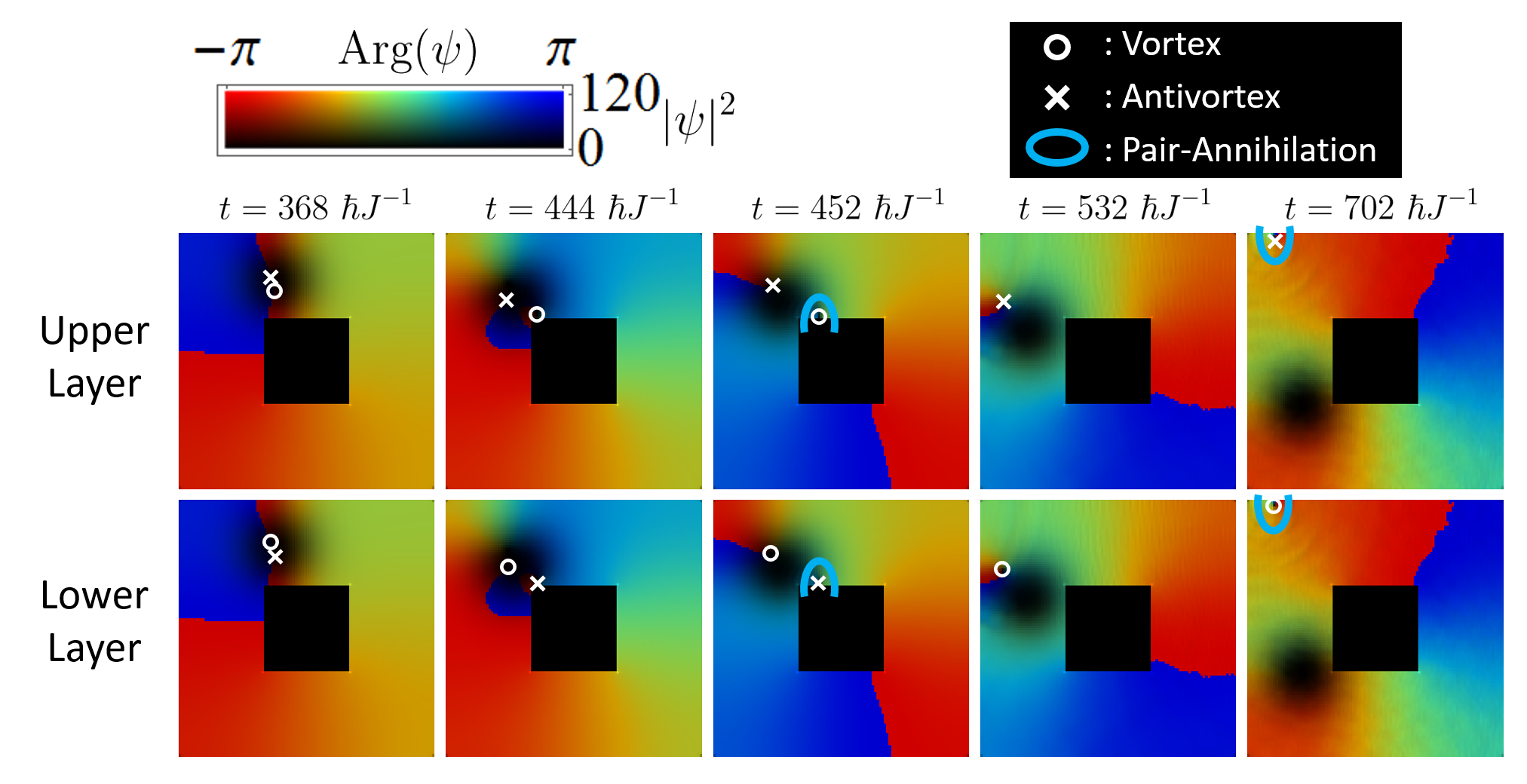}
\caption{Condensate wavefunction during the stirring procedure presented in the leftmost graph in \Cref{stirring_numeric_b}. Determination of vortex and antivortex is opposite in upper and lower layers due to their opposite orientations of surfaces.}\label{stirprocess}
\end{figure}

By observing the course of dynamics of the atomic condensate more closely, one can find that each addition of vorticity is accompanied by a particular procedure of creating, moving, and annihilating vortex-antivortex pairs. For example, the procedure of vorticity addition of the leftmost graph in \Cref{stirring_numeric_b} is illustrated in \Cref{stirprocess}. As shown in the figure, once vortex-antivortex pairs are created on the upper and lower layers, the vortex (antivortex) in the upper (lower) layer moves toward the inner edge, while the antivortex (vortex) in the upper (lower) layer moves toward the outer edge. Around each edge, newly paired vortex and antivortex annihilate with each other. This is topologically equivalent to the operation that moves an antivortex along the loop 2 once, which results in the addition of a unit vorticity to the loop 1. Similarly, an operation that moves an antivortex along the loop 1 once adds a unit vorticity to the loop 2.

\section{Laser-Assisted Tunneling Terms for Quantum Hall Hamiltonian}
Here we introduce the expression of laser-assisted tunneling terms based on the approach of \cite{miyake2013realizing}. To adopt the laser-assisted tunneling in the horizontal direction, we may apply a magnetic field with linear tilts in $x$ and $y$ direction to obtain additional potential $\Delta_x (x/a_x) + \Delta_y (y/a_y)$. We set $\Delta_y \ne \Delta_x$. For the tight-binding model with Wannier basis $\lbrace \ket{w_{n m}} \rbrace$ and bare tunneling strength $J$, overall Hamiltonian with the additional potential is
\begin{eqnarray}
H_0 = \sum_{n,m\in\mathbb{Z}} \left\lbrace (n \Delta_x + m \Delta_y)\ket{w_{n m}}\bra{w_{n m}} - J\left( \ket{w_{(n+1) m}}\bra{w_{n m}} + \ket{w_{n (m+1)}}\bra{w_{n m}} + \text{H.c.} \right) \right\rbrace.
\end{eqnarray}
This Hamiltonian can be diagonalized with Wannier-Stark basis $\lbrace \ket{n m} \rbrace$, which is described by \cite{gluck2002wannier}
\begin{eqnarray}
\ket{n m} = \sum_{r,s\in\mathbb{Z}} J_{r}\left(\frac{2J}{\Delta_x}\right) J_{s}\left(\frac{2J}{\Delta_y}\right) \ket{w_{(n+r)(m+s)}},
\end{eqnarray}
where $J_r$ is the Bessel function of the 1st kind with order $r$. It is straightforward to show that $H_0 \ket{n m} = (n \Delta_x + m \Delta_y)\ket{n m}$ with the aid of recurrence relation of Bessel function, $J_{r-1}(x)+J_{r+1}(x) = 2rJ_{r}(x)/x$.

Now we apply pairs of Raman beams $\mathbf{p}_1$ and $\mathbf{p}_2$ with detuning $c(|\mathbf{p}_1|-|\mathbf{p}_2|)= \Delta_x /\hbar$, $\mathbf{q}_1$ and $\mathbf{q}_2$ with detuning $c(|\mathbf{q}_1|-|\mathbf{q}_2|)= \Delta_y /\hbar$. Then two-photon process between the two beams generate a time-varying dipole potential $V_{\text{dip}}(x,y,t) = \Omega_x \cos(\delta\mathbf{p}\cdot\mathbf{r}-\Delta_x t /\hbar -\theta_x) + \Omega_y \cos(\delta\mathbf{q}\cdot\mathbf{r}-\Delta_y t /\hbar -\theta_y)$, where $\delta\mathbf{p} = \mathbf{p}_1 - \mathbf{p}_2 = p_x \mathbf{\hat{x}} + p_y \mathbf{\hat{y}}$, $\delta\mathbf{q} = \mathbf{q}_1 - \mathbf{q}_2 = q_x \mathbf{\hat{x}} + q_y \mathbf{\hat{y}}$ are relative wave vectors and $\theta_x, \theta_y$ are relative phases between the two beams in each pair. In Wannier-Stark basis in the tight binding limit ($J \ll \Delta_x,\Delta_y$), this dipole potential has following relevant components:
\begin{eqnarray}
\frac{\Omega_x}{2}\braket{n m| e^{i (\delta\mathbf{p}\cdot\mathbf{r}-\Delta_x t /\hbar - \theta_x)}|n m} &=& \frac{\Omega_x}{2} e^{i\theta_p(n,m)} e^{-i(\Delta_x t /\hbar + \theta_x)} + O\left( \frac{\Omega J^2}{\Delta^2}\right) \equiv A^{(x)}_{n m}(t), \nonumber\\
\frac{\Omega_y}{2}\braket{n m|e^{i (\delta\mathbf{q}\cdot\mathbf{r}-\Delta_y t /\hbar - \theta_y)}|n m} &=& \frac{\Omega_y}{2} e^{i\theta_q(n,m)} e^{-i(\Delta_y t /\hbar + \theta_y)} + O\left( \frac{\Omega J^2}{\Delta^2}\right) \equiv A^{(y)}_{n m}(t), \nonumber\\
\frac{\Omega_x}{2} \braket{(n+1) m|e^{\pm i (\delta\mathbf{p}\cdot\mathbf{r}-\Delta_x t /\hbar - \theta_x)}|n m} &=& \frac{\Omega_x J}{2\Delta_x} e^{\pm i\theta_p(n,m)} e^{\mp i (\Delta_x t /\hbar + \theta_x)}\left(1-e^{\pm i p_x a_x}\right) + O\left( \frac{\Omega J^2}{\Delta^2}\right), \nonumber\\
\frac{\Omega_y}{2} \braket{(n+1) m|e^{\pm i (\delta\mathbf{q}\cdot\mathbf{r}-\Delta_y t /\hbar - \theta_y)}|n m} &=& \frac{\Omega_y J}{2\Delta_x} e^{\pm i\theta_q(n,m)} e^{\mp i (\Delta_y t /\hbar + \theta_y)}\left(1-e^{\pm i q_x a_x}\right) + O\left( \frac{\Omega J^2}{\Delta^2}\right), \nonumber\\
\frac{\Omega_x}{2} \braket{n (m+1)|e^{\pm i (\delta\mathbf{p}\cdot\mathbf{r}-\Delta_x t /\hbar - \theta_x)}|n m} &=& \frac{\Omega_x J}{2\Delta_y} e^{\pm i\theta_p(n,m)} e^{\mp i (\Delta_x t /\hbar + \theta_x)}\left(1-e^{\pm i p_y a_y}\right) + O\left( \frac{\Omega J^2}{\Delta^2}\right), \nonumber\\
\frac{\Omega_y}{2} \braket{n (m+1)|e^{\pm i (\delta\mathbf{q}\cdot\mathbf{r}-\Delta_y t /\hbar - \theta_y)}|n m} &=& \frac{\Omega_y J}{2\Delta_y} e^{\pm i\theta_q(n,m)} e^{\mp i (\Delta_y t /\hbar + \theta_y)}\left(1-e^{\pm i q_y a_y}\right) + O\left( \frac{\Omega J^2}{\Delta^2}\right).
\end{eqnarray}
Here, $\theta_p(n,m)\equiv n p_x a_x + m p_y a_y$, $\theta_q(n,m)\equiv n q_x a_x + m q_y a_y$. Now the overall Hamiltonian is
\begin{eqnarray}
H(t) = H_0 + V_{\text{dip}}(t) &=& \sum_{n,m} \left\lbrace \left(n \Delta_x + m \Delta_y + A^{(x)}_{n m}(t) + A^{(x)*}_{n m}(t) + A^{(y)}_{n m}(t) + A^{(y)*}_{n m}(t) \right) \ket{n m}\bra{n m} \right. \nonumber\\
&& \ +\frac{J}{\Delta_x} \left( (1-e^{i p_x a_x})A^{(x)}_{n m}(t) + (1-e^{i q_x a_x})A^{(y)}_{n m}(t) + \text{c.c.} \right) \ket{(n+1) m}\bra{n m} + \text{ H.c.} \nonumber\\
&& \ \left. +\frac{J}{\Delta_y} \left( (1-e^{i p_y a_y})A^{(x)}_{n m}(t) +(1-e^{i q_y a_y}) A^{(y)}_{n m}(t) + \text{c.c.} \right) \ket{n (m+1)}\bra{n m} + \text{ H.c.}\right\rbrace.
\end{eqnarray}

To get rid of differences in diagonal terms, we can use a transformation $U$ to a rotating frame,
\begin{eqnarray}
U = \exp\left[ i\sum_{n,m} \left\lbrace \frac{n \Delta_x + m \Delta_y}{\hbar}t -2 \ \text{Im}\left(\frac{A^{(x)}_{n m}(t)}{\Delta_x}+\frac{A^{(y)}_{n m}(t)}{\Delta_y}\right) \right\rbrace \ket{n m}\bra{n m} \right] \equiv \sum_{n,m} e^{iB(n,m)}\ket{n m}\bra{n m}.
\end{eqnarray}
In this rotating frame, the effective Hamiltonian is given by
\begin{eqnarray}\label{exactHeff}
H_{\text{rot}} &=& UHU^\dag + i\hbar \left(\partial_t U\right) U^\dag
\equiv \sum_{n_x,n_y} J^{(x)}_{n m}(t)\ket{(n+1) m}\bra{n m} + J^{(y)}_{n m}(t)\ket{n (m+1)}\bra{n m} + \text{ H.c.} \nonumber\\
&=& \sum_{n,m} \left\lbrace e^{i (B(n+1,m)-B(n,m))}\frac{J}{\Delta_x} \left( (1-e^{i p_x a_x})A^{(x)}_{n m}(t) + (1-e^{i q_x a_x})A^{(y)}_{n m}(t) + \text{c.c.} \right) \ket{(n+1) m}\bra{n m} + \text{ H.c.} \right. \nonumber\\
&&\left. + e^{i (B(n,m+1)-B(n,m))}\frac{J}{\Delta_y} \left( (1-e^{i p_y a_y})A^{(x)}_{n m}(t) + (1-e^{i q_y a_y})A^{(y)}_{n m}(t) + \text{c.c.} \right) \ket{n (m+1)}\bra{n m} + \text{ H.c.} \right\rbrace.
\end{eqnarray}
By using Jacobi-Anger identity $e^{iz\cos\theta} = \sum_{r\in\mathbb{Z}} i^r J_r(z)e^{ir\theta}$, we get
\begin{eqnarray}
e^{i (B(n+1,m)-B(n,m))} &=& e^{i\Delta_x t/\hbar}
\exp\lbrace -i(2\Omega_x/\Delta_x)\sin(p_x a_x/2)\cos(\theta_p(n,m)+p_x a_x/2-\Delta_x t/\hbar -\theta_x)\rbrace \nonumber\\
&& \times \exp\lbrace -i(2\Omega_y/\Delta_y)\sin(q_x a_x/2)\cos(\theta_q(n,m)+q_x a_x/2-\Delta_y t/\hbar -\theta_y)\rbrace \nonumber\\
&=& e^{i\Delta_x t/\hbar} \sum_{r} J_{r}\left(-\frac{2\Omega_x}{\Delta_x}\sin\left(\frac{p_x a_x}{2}\right)\right) i^{r} \exp\left\lbrace i r(\theta_p(n,m)+p_x a_x/2-\Delta_x t/\hbar -\theta_x) \right\rbrace \nonumber\\
&& \times \sum_{s} J_{s}\left(-\frac{2\Omega_y}{\Delta_y}\sin\left(\frac{q_x a_x}{2}\right)\right) i^{s} \exp\left\lbrace i s(\theta_q(n,m)+q_x a_x/2-\Delta_y t/\hbar -\theta_y) \right\rbrace, \nonumber\\
e^{i (B(n,m+1)-B(n,m))} &=& e^{i\Delta_y t/\hbar}
\exp\lbrace -i(2\Omega_x/\Delta_x)\sin(p_y a_y/2)\cos(\theta_p(n,m)+p_y a_y/2-\Delta_x t/\hbar -\theta_x)\rbrace \nonumber\\
&& \times \exp\lbrace -i(2\Omega_y/\Delta_y)\sin(q_y a_y/2)\cos(\theta_q(n,m)+q_y a_y/2-\Delta_y t/\hbar -\theta_y)\rbrace \nonumber\\
&=& e^{i\Delta_y t/\hbar} \sum_{r} J_{r}\left(-\frac{2\Omega_x}{\Delta_x}\sin\left(\frac{p_y a_y}{2}\right)\right) i^{r} \exp\left\lbrace i r(\theta_p(n,m)+p_y a_y/2-\Delta_x t/\hbar -\theta_x) \right\rbrace \nonumber\\
&& \times \sum_{s} J_{s}\left(-\frac{2\Omega_y}{\Delta_y}\sin\left(\frac{q_y a_y}{2}\right)\right) i^{s} \exp\left\lbrace i s(\theta_q(n,m)+q_y a_y/2-\Delta_y t/\hbar -\theta_y) \right\rbrace.
\end{eqnarray}
For brevity, we define $C_{p,x(y)} \equiv (2\Omega_{x(y)}/\Delta_{x(y)})\sin(p_{x(y)} a_{x(y)}/2)$, $C_{q,x(y)} \equiv (2\Omega_{x(y)}/\Delta_{x(y)})\sin(q_{x(y)} a_{x(y)}/2)$. By time averaging \eqnref{exactHeff} over the time scale $\sim \hbar/\Delta$, we obtain following effective tunneling amplitudes:
\begin{eqnarray}\label{Jeff}
J^{(x)}_{n m ,\text{eff}} &=& \frac{J \Omega_x}{2\Delta_x} \left\lbrace e^{i(\theta_p(n,m)-\theta_x)}(1-e^{i p_x a_x}) J_0\left( C_{p,x} \right) \right. \nonumber\\
&&\qquad \left. - e^{-i(\theta_p(n,m)-\theta_x)}(1-e^{-i p_x a_x}) J_2\left( C_{p,x} \right) e^{i2(\theta_p(n,m)+p_x a_x/2-\theta_x)}
\right\rbrace J_0\left( C_{q,x} \right) \nonumber\\
&=& \frac{J \Omega_x}{2\Delta_x} e^{i(\theta_p(n,m)-\theta_x)}(1-e^{i p_x a_x}) \left\lbrace J_0\left( C_{p,x} \right) + J_2\left( C_{p,x} \right)\right\rbrace J_0\left( C_{q,x} \right) \nonumber\\
&=& J J_1\left( C_{p,x} \right) J_0\left( C_{q,x} \right) \exp\lbrace i(\theta_p(n,m)-\theta_x + (p_x a_x -\pi)/2)\rbrace, \nonumber\\
J^{(y)}_{n m ,\text{eff}} &=& \frac{J \Omega_y}{2\Delta_y} \left\lbrace e^{i(\theta_q(n,m)-\theta_y)}(1-e^{i q_y a_y}) J_0\left( C_{q,y} \right) \right. \nonumber\\
&&\qquad \left. - e^{-i(\theta_q(n,m)-\theta_y)}(1-e^{-i q_y a_y}) J_2\left( C_{q,y} \right) e^{i2(\theta_q(n,m)+p_y a_y/2-\theta_y)}
\right\rbrace J_0\left( C_{p,y} \right) \nonumber\\
&=& \frac{J \Omega_y}{2\Delta_y} e^{i(\theta_q(n,m)-\theta_y)}(1-e^{i q_y a_y}) \left\lbrace J_0\left( C_{q,y} \right) + J_2\left( C_{q,y} \right)\right\rbrace J_0\left( C_{p,y} \right) \nonumber\\
&=& J J_1\left( C_{q,y} \right) J_0\left( C_{p,y} \right) \exp\lbrace i(\theta_q(n,m)-\theta_y + (q_y a_y -\pi)/2)\rbrace.
\end{eqnarray}

Since $\theta_p$ and $\theta_q$ are linear in $n$ and $m$, the resulting effective Hamiltonian describes the charged particle under the presence of a uniform magnetic field \cite{hofstadter1976energy}. These expressions can be further simplified in perturbative regime, $\Omega_x, \Omega_y \ll \Delta_x, \Delta_y$. In such case, $C_{p,x(y)}, C_{q,x(y)} \ll 1$, so $J_0(C) = 1 + O(C^2)$ and $J_1(C) = C/2 + O(C^2)$. Then by setting $\theta_x = (p_x a_x + \pi)/2$ and $\theta_y = (q_y a_y + \pi)/2$, we get
\begin{eqnarray}\label{JeffPerturb}
J^{(x)}_{n m ,\text{eff}} &=& -\frac{J \Omega_x}{\Delta_x} \sin\left(\frac{p_x a_x}{2}\right) \exp\lbrace i(n p_x a_x + m p_y a_y)\rbrace, \nonumber\\
J^{(y)}_{n m ,\text{eff}} &=& -\frac{J \Omega_y}{\Delta_y} \sin\left(\frac{q_y a_y}{2}\right) \exp\lbrace i(n q_x a_x + m q_y a_y)\rbrace.
\end{eqnarray}

\section{Beam Configuration for Quantum Hall Hamiltonian on Torus}

For the simplicity of construction, we assume our lattice spans from $(n,m)=(1,1)$ to $(n,m)=(L,L)=(p+2q+1,p+2q+1)$. Here, $p$ is the width of the square puncture in the middle while $q$ is the width of the square annulus, in the unit of lattice spacing. To obtain the tunneling phases shown in \eqnref{fqh_hamiltonian}, we need our Raman beams to satisfy following conditions:
\begin{eqnarray}\label{cond_interfere}
(\mathbf{k}_{i+})_z = - (\mathbf{k}_{i-})_z = (\pi/4) z_0^{-1}, \quad i = 1 \text{ to } 4.
\end{eqnarray}
\begin{eqnarray}\label{cond_detuning}
& |\mathbf{k}_1| - |\mathbf{k}_{1\pm}|  = |\mathbf{k}_{3\pm}| - |\mathbf{k}_3| 
= (\omega_1 - \omega_{1\pm})/c = (\omega_{3\pm} - \omega_3)/c = \Delta_x /\hbar c, \nonumber\\ 
& |\mathbf{k}_2| - |\mathbf{k}_{2\pm}|  = |\mathbf{k}_{4\pm}| - |\mathbf{k}_4|
= (\omega_2 - \omega_{2\pm})/c = (\omega_{4\pm} - \omega_4)/c = \Delta_y /\hbar c.
\end{eqnarray}
\begin{eqnarray}\label{cond_phase}
& (\mathbf{k}_1-\mathbf{k}_{1\pm})_x = (\mathbf{k}_2-\mathbf{k}_{2\pm})_x =
(\mathbf{k}_{3\pm}-\mathbf{k}_3)_x = (\mathbf{k}_4-\mathbf{k}_{4\pm})_x = \phi/2a_x, \nonumber\\
& (\mathbf{k}_{1\pm}-\mathbf{k}_1)_y = (\mathbf{k}_{2\pm}-\mathbf{k}_2)_y =
(\mathbf{k}_{3\pm}-\mathbf{k}_3)_y = (\mathbf{k}_4-\mathbf{k}_{4\pm})_y = \phi/2a_y.
\end{eqnarray}
\begin{eqnarray}\label{cond_commensurate}
(p+q)\phi \text{ mod } 2\pi = (q+2)\phi \text{ mod } 2\pi = 0.
\end{eqnarray}
Here, the relative phase between beams $\mathbf{k}_{i+}$ and $\mathbf{k}_{i-}$ are adjusted in a way that the vertical standing wave between them to have a node at $z = -z_0$ ($i=1,2$) or $z = z_0$ ($i=3,4$). That is, the beam triplets $T_{\mathbf{1}}$ and $T_{\mathbf{2}}$ target the upper layer while the beam triplets $T_{\mathbf{3}}$ and $T_{\mathbf{4}}$ target the lower layer. \eqnref{cond_interfere} is required to guarantee that beams labeled with $+$ and beams labeled with $-$ to show constructive interference at targeted layer while they destructively interfere at the non-targeted layer. \eqnref{cond_detuning} implies that $T_{\mathbf{1}}$ and $T_{\mathbf{3}}$ give laser-assisted tunneling in the $x$ direction while $T_{\mathbf{2}}$ and $T_{\mathbf{4}}$ give laser-assisted tunneling in the $y$ direction. \eqnref{cond_phase} ensures that synthetic magnetic flux threading each plaquette in the outward direction to be $\phi$, with a choice of symmetric gauge. Expressions for laser-assisted tunneling terms are identified in \eqnref{Jeff} and \eqnref{JeffPerturb}. With given conditions, phases of the tunneling terms in \eqnref{JeffPerturb} are given by
\begin{eqnarray}
\text{Arg}\left(-J^{(x)}_{n m ,\text{eff}}\right) &=& \left\lbrace
\begin{array}{cc}
(\mathbf{k_1}-\mathbf{k_{1,\pm}})\cdot (n a_x \mathbf{\hat{x}} + m a_x \mathbf{\hat{x}}) = (n -m)\phi/2
& \quad\text{upper layer} \\
(\mathbf{k_{3,\pm}}-\mathbf{k_3})\cdot (n a_x \mathbf{\hat{x}} + m a_x \mathbf{\hat{x}}) = (n +m)\phi/2
& \quad\text{lower layer}
\end{array} \right. ,\nonumber\\
\text{Arg}\left(-J^{(y)}_{n m ,\text{eff}}\right) &=& \left\lbrace
\begin{array}{cc}
(\mathbf{k_2}-\mathbf{k_{2,\pm}})\cdot (n a_x \mathbf{\hat{x}} + m a_x \mathbf{\hat{x}}) = (n -m)\phi/2
& \quad\text{upper layer} \\
(\mathbf{k_{4,\pm}}-\mathbf{k_4})\cdot (n a_x \mathbf{\hat{x}} + m a_x \mathbf{\hat{x}}) = -(n +m)\phi/2
& \quad\text{lower layer}
\end{array} \right. .
\end{eqnarray}

To have the uniform synthetic magnetic field all over the torus surface, we need to make every plaquette in the side areas to have flux $\phi$ in the outward direction. Keeping the inter-layer tunneling real, the outward flux from each plaquette in side areas are shown in \Cref{side_a}. While the outward fluxes from different sides are not identical in general, we may set all of them to be identical up to modulo of $2\pi$. That is,
\begin{eqnarray}
-(q+1)\phi \text{ mod } 2\pi = (p+q+1)\phi \text{ mod } 2\pi = -(2q+p+1)\phi \text{ mod } 2\pi = \phi,
\end{eqnarray}
which is equivalent to the condition \eqnref{cond_commensurate}. To illustrate how this scheme works altogether, tunneling phases in a lattice with $p=2,q=4,\phi=\pi/3$ is shown in \Cref{side_b}. From this figure, one can check that outward flux from every single plaquette is the same as $\phi$.

\begin{figure}
\centering
\includegraphics[width=\linewidth]{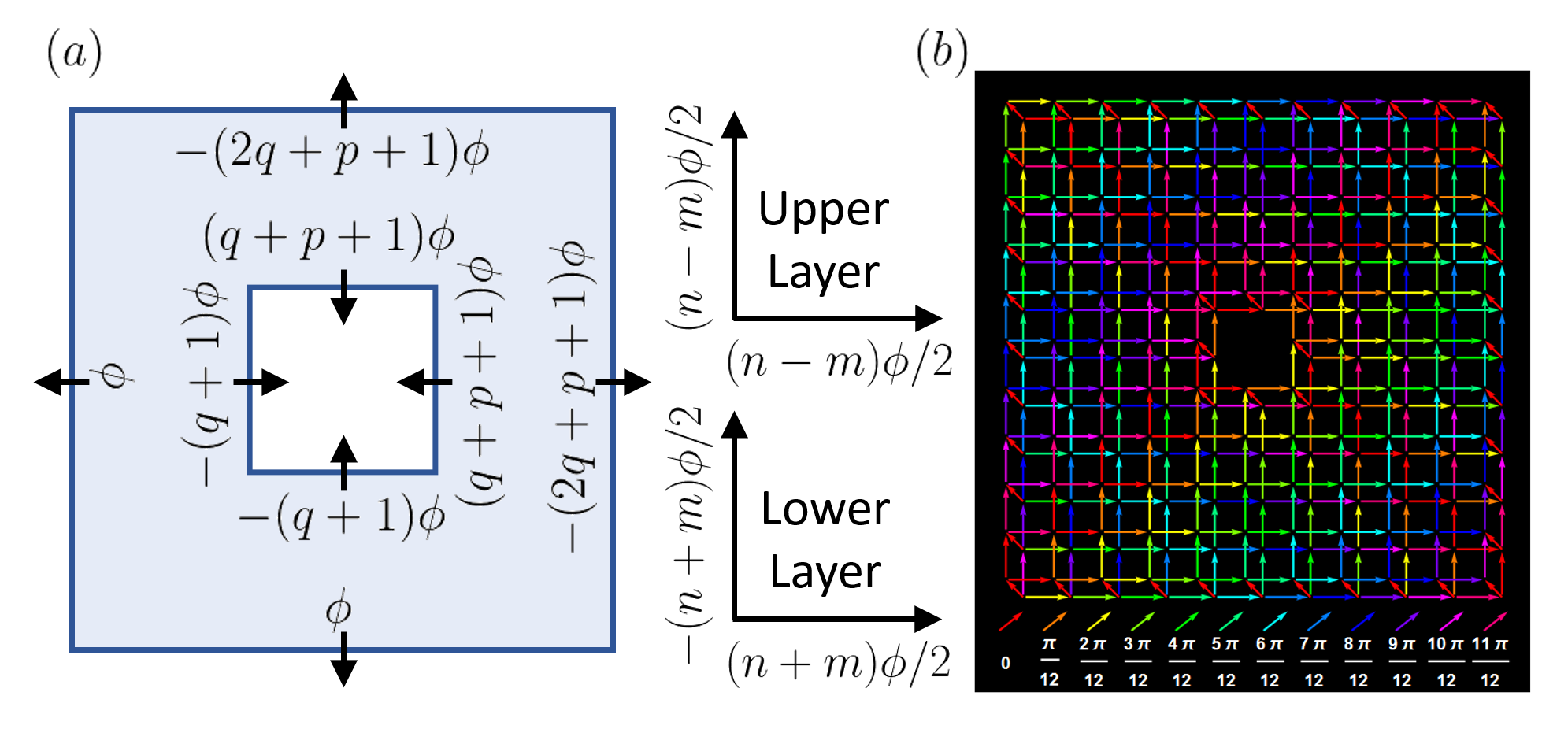}\subfigure{\label{side_a}}\subfigure{\label{side_b}}
\caption{
(a) Flux per plaquette toward the outside of torus along the side areas. $p$ is the width of the square puncture in the middle, and $q$ is the width of the square annulus. Size of the square lattice is then $L \times L$, $L=2q+p+1$.
(b) tunneling phases of sample lattice with $p=2,q=4,\phi=\pi/3$.}\label{sideflux}
\end{figure}

\section{Measurement of Topological Degeneracy}

In \Cref{fqhe_g}, the spectral flow and the flow of the $y$-coordinates in the changing twist angle show very similar graphs to each other. To briefly understand the physics behind this resemblance, we can consider the thin torus limit \cite{bergholtz2005half}. For a short interaction length, the two-fold degenerate ground states for $\nu=1/2$ are given as $\ket{010101\cdots}$ and $\ket{101010\cdots}$ where each 0 and 1 indicates the occupation at each orbital. While there is some freedom to choose these orbitals, we select the orbitals localized in the $y$ direction on the torus $(x,y)\in [0,L_x)\times [0,L_y)$. If a perturbative potential $V(y) = (\Delta V_y /L_y)y$ is applied, the energy shift in each ground state is proportional to the $y$-coordinate expectation values of each state, as shown in \Cref{EY_a}. Since this energy splitting is a finite-size effect which should vanish in the thermodynamic limit, the proportionality constant presented in this thin torus limit is not precise. Yet, in a finite-size system, this effect can be experimentally detected.

\begin{figure}[h]
\centering
\includegraphics[width=\linewidth]{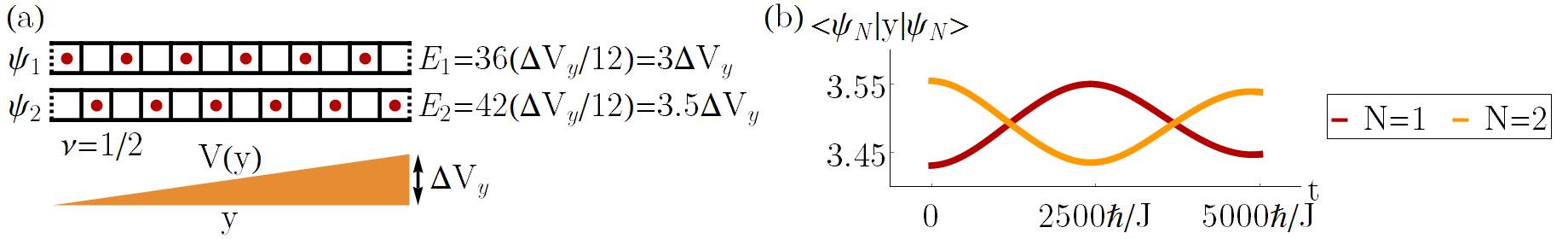}\subfigure{\label{EY_a}}\subfigure{\label{EY_b}}
\caption{(a) The relation between the energy splitting and difference in $y$-coordinates becomes clear in the thin torus limit. Case of $N_p=6, \nu=1/2$ is shown for an example. (b) Average $y$-coordinates during the adiabatic flux insertion, in the presence of potential $V(y) = (0.01J/L_y)y$. Flux corresponds to $\alpha_x=4\pi$ is inserted over the time interval of $5000\hbar/J$.}\label{EYsplitting}
\end{figure}

With a proper cooling scheme, we can prepare atoms to be in a particular ground state due to the energy splitting. Yet, to observe the $4\pi$-periodicity in spectral flow of each state, the flux should to be inserted adiabatically. To see if such an adibatic evolution is possible, we simulate the procedure of flux insertion on the system in \Cref{fqhe_g} [\Cref{EY_b}]. As shown in the figure, change in $y$-coordinates can be detected in a proper adiabatic time evolution. By measuring the atomic densities for the varying twist angle, one can detect the anticipated periodicity and therefore measure the topological degeneracy.

\end{widetext}
\end{appendix}

%\bibliography{biblio}

\begin{thebibliography}{44}%
\makeatletter
\providecommand \@ifxundefined [1]{%
 \@ifx{#1\undefined}
}%
\providecommand \@ifnum [1]{%
 \ifnum #1\expandafter \@firstoftwo
 \else \expandafter \@secondoftwo
 \fi
}%
\providecommand \@ifx [1]{%
 \ifx #1\expandafter \@firstoftwo
 \else \expandafter \@secondoftwo
 \fi
}%
\providecommand \natexlab [1]{#1}%
\providecommand \enquote  [1]{``#1''}%
\providecommand \bibnamefont  [1]{#1}%
\providecommand \bibfnamefont [1]{#1}%
\providecommand \citenamefont [1]{#1}%
\providecommand \href@noop [0]{\@secondoftwo}%
\providecommand \href [0]{\begingroup \@sanitize@url \@href}%
\providecommand \@href[1]{\@@startlink{#1}\@@href}%
\providecommand \@@href[1]{\endgroup#1\@@endlink}%
\providecommand \@sanitize@url [0]{\catcode `\\12\catcode `\$12\catcode
  `\&12\catcode `\#12\catcode `\^12\catcode `\_12\catcode `\%12\relax}%
\providecommand \@@startlink[1]{}%
\providecommand \@@endlink[0]{}%
\providecommand \url  [0]{\begingroup\@sanitize@url \@url }%
\providecommand \@url [1]{\endgroup\@href {#1}{\urlprefix }}%
\providecommand \urlprefix  [0]{URL }%
\providecommand \Eprint [0]{\href }%
\providecommand \doibase [0]{http://dx.doi.org/}%
\providecommand \selectlanguage [0]{\@gobble}%
\providecommand \bibinfo  [0]{\@secondoftwo}%
\providecommand \bibfield  [0]{\@secondoftwo}%
\providecommand \translation [1]{[#1]}%
\providecommand \BibitemOpen [0]{}%
\providecommand \bibitemStop [0]{}%
\providecommand \bibitemNoStop [0]{.\EOS\space}%
\providecommand \EOS [0]{\spacefactor3000\relax}%
\providecommand \BibitemShut  [1]{\csname bibitem#1\endcsname}%
\let\auto@bib@innerbib\@empty
%</preamble>
\bibitem [{\citenamefont {Jaksch}\ \emph {et~al.}(1998)\citenamefont {Jaksch},
  \citenamefont {Bruder}, \citenamefont {Cirac}, \citenamefont {Gardiner},\
  and\ \citenamefont {Zoller}}]{jaksch1998cold}%
  \BibitemOpen
  \bibfield  {author} {\bibinfo {author} {\bibfnamefont {D.}~\bibnamefont
  {Jaksch}}, \bibinfo {author} {\bibfnamefont {C.}~\bibnamefont {Bruder}},
  \bibinfo {author} {\bibfnamefont {J.~I.}\ \bibnamefont {Cirac}}, \bibinfo
  {author} {\bibfnamefont {C.~W.}\ \bibnamefont {Gardiner}}, \ and\ \bibinfo
  {author} {\bibfnamefont {P.}~\bibnamefont {Zoller}},\ }\href
  {https://journals.aps.org/prl/abstract/10.1103/PhysRevLett.81.3108}
  {\bibfield  {journal} {\bibinfo  {journal} {Phys. Rev. Lett.}\ }\textbf
  {\bibinfo {volume} {81}},\ \bibinfo {pages} {3108} (\bibinfo {year}
  {1998})}\BibitemShut {NoStop}%
\bibitem [{\citenamefont {Greiner}\ \emph {et~al.}(2002)\citenamefont
  {Greiner}, \citenamefont {Mandel}, \citenamefont {Esslinger}, \citenamefont
  {H{\"a}nsch},\ and\ \citenamefont {Bloch}}]{greiner2002quantum}%
  \BibitemOpen
  \bibfield  {author} {\bibinfo {author} {\bibfnamefont {M.}~\bibnamefont
  {Greiner}}, \bibinfo {author} {\bibfnamefont {O.}~\bibnamefont {Mandel}},
  \bibinfo {author} {\bibfnamefont {T.}~\bibnamefont {Esslinger}}, \bibinfo
  {author} {\bibfnamefont {T.~W.}\ \bibnamefont {H{\"a}nsch}}, \ and\ \bibinfo
  {author} {\bibfnamefont {I.}~\bibnamefont {Bloch}},\ }\href
  {https://www.nature.com/articles/415039a} {\bibfield  {journal} {\bibinfo
  {journal} {Nature (London)}\ }\textbf {\bibinfo {volume} {415}},\ \bibinfo
  {pages} {39} (\bibinfo {year} {2002})}\BibitemShut {NoStop}%
\bibitem [{\citenamefont {Aidelsburger}\ \emph {et~al.}(2013)\citenamefont
  {Aidelsburger}, \citenamefont {Atala}, \citenamefont {Lohse}, \citenamefont
  {Barreiro}, \citenamefont {Paredes},\ and\ \citenamefont
  {Bloch}}]{aidelsburger2013realization}%
  \BibitemOpen
  \bibfield  {author} {\bibinfo {author} {\bibfnamefont {M.}~\bibnamefont
  {Aidelsburger}}, \bibinfo {author} {\bibfnamefont {M.}~\bibnamefont {Atala}},
  \bibinfo {author} {\bibfnamefont {M.}~\bibnamefont {Lohse}}, \bibinfo
  {author} {\bibfnamefont {J.~T.}\ \bibnamefont {Barreiro}}, \bibinfo {author}
  {\bibfnamefont {B.}~\bibnamefont {Paredes}}, \ and\ \bibinfo {author}
  {\bibfnamefont {I.}~\bibnamefont {Bloch}},\ }\href
  {https://journals.aps.org/prl/abstract/10.1103/PhysRevLett.111.185301}
  {\bibfield  {journal} {\bibinfo  {journal} {Phys. Rev. Lett.}\ }\textbf
  {\bibinfo {volume} {111}},\ \bibinfo {pages} {185301} (\bibinfo {year}
  {2013})}\BibitemShut {NoStop}%
\bibitem [{\citenamefont {Miyake}\ \emph {et~al.}(2013)\citenamefont {Miyake},
  \citenamefont {Siviloglou}, \citenamefont {Kennedy}, \citenamefont {Burton},\
  and\ \citenamefont {Ketterle}}]{miyake2013realizing}%
  \BibitemOpen
  \bibfield  {author} {\bibinfo {author} {\bibfnamefont {H.}~\bibnamefont
  {Miyake}}, \bibinfo {author} {\bibfnamefont {G.~A.}\ \bibnamefont
  {Siviloglou}}, \bibinfo {author} {\bibfnamefont {C.~J.}\ \bibnamefont
  {Kennedy}}, \bibinfo {author} {\bibfnamefont {W.~C.}\ \bibnamefont {Burton}},
  \ and\ \bibinfo {author} {\bibfnamefont {W.}~\bibnamefont {Ketterle}},\
  }\href {https://journals.aps.org/prl/abstract/10.1103/PhysRevLett.111.185302}
  {\bibfield  {journal} {\bibinfo  {journal} {Phys. Rev. Lett.}\ }\textbf
  {\bibinfo {volume} {111}},\ \bibinfo {pages} {185302} (\bibinfo {year}
  {2013})}\BibitemShut {NoStop}%
\bibitem [{\citenamefont {Paredes}\ \emph {et~al.}(2004)\citenamefont
  {Paredes}, \citenamefont {Widera}, \citenamefont {Murg}, \citenamefont
  {Mandel}, \citenamefont {F{\"o}lling}, \citenamefont {Cirac}, \citenamefont
  {Shlyapnikov}, \citenamefont {H{\"a}nsch},\ and\ \citenamefont
  {Bloch}}]{paredes2004tonks}%
  \BibitemOpen
  \bibfield  {author} {\bibinfo {author} {\bibfnamefont {B.}~\bibnamefont
  {Paredes}}, \bibinfo {author} {\bibfnamefont {A.}~\bibnamefont {Widera}},
  \bibinfo {author} {\bibfnamefont {V.}~\bibnamefont {Murg}}, \bibinfo {author}
  {\bibfnamefont {O.}~\bibnamefont {Mandel}}, \bibinfo {author} {\bibfnamefont
  {S.}~\bibnamefont {F{\"o}lling}}, \bibinfo {author} {\bibfnamefont
  {I.}~\bibnamefont {Cirac}}, \bibinfo {author} {\bibfnamefont {G.~V.}\
  \bibnamefont {Shlyapnikov}}, \bibinfo {author} {\bibfnamefont {T.~W.}\
  \bibnamefont {H{\"a}nsch}}, \ and\ \bibinfo {author} {\bibfnamefont
  {I.}~\bibnamefont {Bloch}},\ }\href
  {https://www.nature.com/articles/nature02530} {\bibfield  {journal} {\bibinfo
   {journal} {Nature (London)}\ }\textbf {\bibinfo {volume} {429}},\ \bibinfo
  {pages} {277} (\bibinfo {year} {2004})}\BibitemShut {NoStop}%
\bibitem [{\citenamefont {Greiner}\ \emph {et~al.}(2001)\citenamefont
  {Greiner}, \citenamefont {Bloch}, \citenamefont {Mandel}, \citenamefont
  {H{\"a}nsch},\ and\ \citenamefont {Esslinger}}]{greiner2001bose}%
  \BibitemOpen
  \bibfield  {author} {\bibinfo {author} {\bibfnamefont {M.}~\bibnamefont
  {Greiner}}, \bibinfo {author} {\bibfnamefont {I.}~\bibnamefont {Bloch}},
  \bibinfo {author} {\bibfnamefont {O.}~\bibnamefont {Mandel}}, \bibinfo
  {author} {\bibfnamefont {T.}~\bibnamefont {H{\"a}nsch}}, \ and\ \bibinfo
  {author} {\bibfnamefont {T.}~\bibnamefont {Esslinger}},\ }\href
  {https://link.springer.com/article/10.1007/s003400100744} {\bibfield
  {journal} {\bibinfo  {journal} {Appl. Phys. B: Lasers and Optics}\ }\textbf
  {\bibinfo {volume} {73}},\ \bibinfo {pages} {769} (\bibinfo {year}
  {2001})}\BibitemShut {NoStop}%
\bibitem [{\citenamefont {St{\"o}ferle}\ \emph {et~al.}(2004)\citenamefont
  {St{\"o}ferle}, \citenamefont {Moritz}, \citenamefont {Schori}, \citenamefont
  {K{\"o}hl},\ and\ \citenamefont {Esslinger}}]{stoferle2004transition}%
  \BibitemOpen
  \bibfield  {author} {\bibinfo {author} {\bibfnamefont {T.}~\bibnamefont
  {St{\"o}ferle}}, \bibinfo {author} {\bibfnamefont {H.}~\bibnamefont
  {Moritz}}, \bibinfo {author} {\bibfnamefont {C.}~\bibnamefont {Schori}},
  \bibinfo {author} {\bibfnamefont {M.}~\bibnamefont {K{\"o}hl}}, \ and\
  \bibinfo {author} {\bibfnamefont {T.}~\bibnamefont {Esslinger}},\ }\href
  {https://journals.aps.org/prl/abstract/10.1103/PhysRevLett.92.130403}
  {\bibfield  {journal} {\bibinfo  {journal} {Phys. Rev. Lett.}\ }\textbf
  {\bibinfo {volume} {92}},\ \bibinfo {pages} {130403} (\bibinfo {year}
  {2004})}\BibitemShut {NoStop}%
\bibitem [{\citenamefont {Spielman}\ \emph {et~al.}(2007)\citenamefont
  {Spielman}, \citenamefont {Phillips},\ and\ \citenamefont
  {Porto}}]{spielman2007mott}%
  \BibitemOpen
  \bibfield  {author} {\bibinfo {author} {\bibfnamefont {I.~B.}\ \bibnamefont
  {Spielman}}, \bibinfo {author} {\bibfnamefont {W.~D.}\ \bibnamefont
  {Phillips}}, \ and\ \bibinfo {author} {\bibfnamefont {J.~V.}\ \bibnamefont
  {Porto}},\ }\href
  {https://journals.aps.org/prl/abstract/10.1103/PhysRevLett.98.080404}
  {\bibfield  {journal} {\bibinfo  {journal} {Phys. Rev. Lett.}\ }\textbf
  {\bibinfo {volume} {98}},\ \bibinfo {pages} {080404} (\bibinfo {year}
  {2007})}\BibitemShut {NoStop}%
\bibitem [{\citenamefont {Becker}\ \emph {et~al.}(2010)\citenamefont {Becker},
  \citenamefont {Soltan-Panahi}, \citenamefont {Kronj{\"a}ger}, \citenamefont
  {D{\"o}rscher}, \citenamefont {Bongs},\ and\ \citenamefont
  {Sengstock}}]{becker2010ultracold}%
  \BibitemOpen
  \bibfield  {author} {\bibinfo {author} {\bibfnamefont {C.}~\bibnamefont
  {Becker}}, \bibinfo {author} {\bibfnamefont {P.}~\bibnamefont
  {Soltan-Panahi}}, \bibinfo {author} {\bibfnamefont {J.}~\bibnamefont
  {Kronj{\"a}ger}}, \bibinfo {author} {\bibfnamefont {S.}~\bibnamefont
  {D{\"o}rscher}}, \bibinfo {author} {\bibfnamefont {K.}~\bibnamefont {Bongs}},
  \ and\ \bibinfo {author} {\bibfnamefont {K.}~\bibnamefont {Sengstock}},\
  }\href {http://iopscience.iop.org/article/10.1088/1367-2630/12/6/065025/meta}
  {\bibfield  {journal} {\bibinfo  {journal} {New J. Phys.}\ }\textbf {\bibinfo
  {volume} {12}},\ \bibinfo {pages} {065025} (\bibinfo {year}
  {2010})}\BibitemShut {NoStop}%
\bibitem [{\citenamefont {Tarruell}\ \emph {et~al.}(2012)\citenamefont
  {Tarruell}, \citenamefont {Greif}, \citenamefont {Uehlinger}, \citenamefont
  {Jotzu},\ and\ \citenamefont {Esslinger}}]{tarruell2012creating}%
  \BibitemOpen
  \bibfield  {author} {\bibinfo {author} {\bibfnamefont {L.}~\bibnamefont
  {Tarruell}}, \bibinfo {author} {\bibfnamefont {D.}~\bibnamefont {Greif}},
  \bibinfo {author} {\bibfnamefont {T.}~\bibnamefont {Uehlinger}}, \bibinfo
  {author} {\bibfnamefont {G.}~\bibnamefont {Jotzu}}, \ and\ \bibinfo {author}
  {\bibfnamefont {T.}~\bibnamefont {Esslinger}},\ }\href
  {https://www.nature.com/articles/nature10871} {\bibfield  {journal} {\bibinfo
   {journal} {Nature (London)}\ }\textbf {\bibinfo {volume} {483}},\ \bibinfo
  {pages} {302} (\bibinfo {year} {2012})}\BibitemShut {NoStop}%
\bibitem [{\citenamefont {Jo}\ \emph {et~al.}(2012)\citenamefont {Jo},
  \citenamefont {Guzman}, \citenamefont {Thomas}, \citenamefont {Hosur},
  \citenamefont {Vishwanath},\ and\ \citenamefont
  {Stamper-Kurn}}]{jo2012ultracold}%
  \BibitemOpen
  \bibfield  {author} {\bibinfo {author} {\bibfnamefont {G.-B.}\ \bibnamefont
  {Jo}}, \bibinfo {author} {\bibfnamefont {J.}~\bibnamefont {Guzman}}, \bibinfo
  {author} {\bibfnamefont {C.~K.}\ \bibnamefont {Thomas}}, \bibinfo {author}
  {\bibfnamefont {P.}~\bibnamefont {Hosur}}, \bibinfo {author} {\bibfnamefont
  {A.}~\bibnamefont {Vishwanath}}, \ and\ \bibinfo {author} {\bibfnamefont
  {D.~M.}\ \bibnamefont {Stamper-Kurn}},\ }\href
  {https://journals.aps.org/prl/abstract/10.1103/PhysRevLett.108.045305}
  {\bibfield  {journal} {\bibinfo  {journal} {Phys. Rev. Lett.}\ }\textbf
  {\bibinfo {volume} {108}},\ \bibinfo {pages} {045305} (\bibinfo {year}
  {2012})}\BibitemShut {NoStop}%
\bibitem [{\citenamefont {Ramanathan}\ \emph {et~al.}(2011)\citenamefont
  {Ramanathan}, \citenamefont {Wright}, \citenamefont {Muniz}, \citenamefont
  {Zelan}, \citenamefont {Hill~III}, \citenamefont {Lobb}, \citenamefont
  {Helmerson}, \citenamefont {Phillips},\ and\ \citenamefont
  {Campbell}}]{ramanathan2011superflow}%
  \BibitemOpen
  \bibfield  {author} {\bibinfo {author} {\bibfnamefont {A.}~\bibnamefont
  {Ramanathan}}, \bibinfo {author} {\bibfnamefont {K.}~\bibnamefont {Wright}},
  \bibinfo {author} {\bibfnamefont {S.}~\bibnamefont {Muniz}}, \bibinfo
  {author} {\bibfnamefont {M.}~\bibnamefont {Zelan}}, \bibinfo {author}
  {\bibfnamefont {W.}~\bibnamefont {Hill~III}}, \bibinfo {author}
  {\bibfnamefont {C.}~\bibnamefont {Lobb}}, \bibinfo {author} {\bibfnamefont
  {K.}~\bibnamefont {Helmerson}}, \bibinfo {author} {\bibfnamefont
  {W.}~\bibnamefont {Phillips}}, \ and\ \bibinfo {author} {\bibfnamefont
  {G.}~\bibnamefont {Campbell}},\ }\href
  {https://journals.aps.org/prl/abstract/10.1103/PhysRevLett.106.130401}
  {\bibfield  {journal} {\bibinfo  {journal} {Phys. Rev. Lett.}\ }\textbf
  {\bibinfo {volume} {106}},\ \bibinfo {pages} {130401} (\bibinfo {year}
  {2011})}\BibitemShut {NoStop}%
\bibitem [{\citenamefont {{\L}{\k{a}}cki}\ \emph {et~al.}(2016)\citenamefont
  {{\L}{\k{a}}cki}, \citenamefont {Pichler}, \citenamefont {Sterdyniak},
  \citenamefont {Lyras}, \citenamefont {Lembessis}, \citenamefont {Al-Dossary},
  \citenamefont {Budich},\ and\ \citenamefont {Zoller}}]{lkacki2016quantum}%
  \BibitemOpen
  \bibfield  {author} {\bibinfo {author} {\bibfnamefont {M.}~\bibnamefont
  {{\L}{\k{a}}cki}}, \bibinfo {author} {\bibfnamefont {H.}~\bibnamefont
  {Pichler}}, \bibinfo {author} {\bibfnamefont {A.}~\bibnamefont {Sterdyniak}},
  \bibinfo {author} {\bibfnamefont {A.}~\bibnamefont {Lyras}}, \bibinfo
  {author} {\bibfnamefont {V.~E.}\ \bibnamefont {Lembessis}}, \bibinfo {author}
  {\bibfnamefont {O.}~\bibnamefont {Al-Dossary}}, \bibinfo {author}
  {\bibfnamefont {J.~C.}\ \bibnamefont {Budich}}, \ and\ \bibinfo {author}
  {\bibfnamefont {P.}~\bibnamefont {Zoller}},\ }\href
  {https://journals.aps.org/pra/abstract/10.1103/PhysRevA.93.013604} {\bibfield
   {journal} {\bibinfo  {journal} {Phys. Rev. A}\ }\textbf {\bibinfo {volume}
  {93}},\ \bibinfo {pages} {013604} (\bibinfo {year} {2016})}\BibitemShut
  {NoStop}%
\bibitem [{\citenamefont {Stuhl}\ \emph {et~al.}(2015)\citenamefont {Stuhl},
  \citenamefont {Lu}, \citenamefont {Aycock}, \citenamefont {Genkina},\ and\
  \citenamefont {Spielman}}]{stuhl2015visualizing}%
  \BibitemOpen
  \bibfield  {author} {\bibinfo {author} {\bibfnamefont {B.}~\bibnamefont
  {Stuhl}}, \bibinfo {author} {\bibfnamefont {H.-I.}\ \bibnamefont {Lu}},
  \bibinfo {author} {\bibfnamefont {L.}~\bibnamefont {Aycock}}, \bibinfo
  {author} {\bibfnamefont {D.}~\bibnamefont {Genkina}}, \ and\ \bibinfo
  {author} {\bibfnamefont {I.}~\bibnamefont {Spielman}},\ }\href
  {http://science.sciencemag.org/content/349/6255/1514} {\bibfield  {journal}
  {\bibinfo  {journal} {Science}\ }\textbf {\bibinfo {volume} {349}},\ \bibinfo
  {pages} {1514} (\bibinfo {year} {2015})}\BibitemShut {NoStop}%
\bibitem [{\citenamefont {Haldane}(1985)}]{haldane1985many}%
  \BibitemOpen
  \bibfield  {author} {\bibinfo {author} {\bibfnamefont {F.}~\bibnamefont
  {Haldane}},\ }\href
  {https://journals.aps.org/prl/abstract/10.1103/PhysRevLett.55.2095}
  {\bibfield  {journal} {\bibinfo  {journal} {Phys. Rev. Lett.}\ }\textbf
  {\bibinfo {volume} {55}},\ \bibinfo {pages} {2095} (\bibinfo {year}
  {1985})}\BibitemShut {NoStop}%
\bibitem [{\citenamefont {Wen}\ and\ \citenamefont
  {Niu}(1990)}]{wen1990ground}%
  \BibitemOpen
  \bibfield  {author} {\bibinfo {author} {\bibfnamefont {X.-G.}\ \bibnamefont
  {Wen}}\ and\ \bibinfo {author} {\bibfnamefont {Q.}~\bibnamefont {Niu}},\
  }\href {https://journals.aps.org/prb/abstract/10.1103/PhysRevB.41.9377}
  {\bibfield  {journal} {\bibinfo  {journal} {Phys. Rev. B}\ }\textbf {\bibinfo
  {volume} {41}},\ \bibinfo {pages} {9377} (\bibinfo {year}
  {1990})}\BibitemShut {NoStop}%
\bibitem [{\citenamefont {Kitaev}(2003)}]{kitaev2003fault}%
  \BibitemOpen
  \bibfield  {author} {\bibinfo {author} {\bibfnamefont {A.~Y.}\ \bibnamefont
  {Kitaev}},\ }\href
  {https://www.sciencedirect.com/science/article/pii/S0003491602000180}
  {\bibfield  {journal} {\bibinfo  {journal} {Ann. Phys. (N.Y.)}\ }\textbf
  {\bibinfo {volume} {303}},\ \bibinfo {pages} {2} (\bibinfo {year}
  {2003})}\BibitemShut {NoStop}%
\bibitem [{\citenamefont {Wen}(2002)}]{wen2002quantum}%
  \BibitemOpen
  \bibfield  {author} {\bibinfo {author} {\bibfnamefont {X.-G.}\ \bibnamefont
  {Wen}},\ }\href
  {https://journals.aps.org/prb/abstract/10.1103/PhysRevB.65.165113} {\bibfield
   {journal} {\bibinfo  {journal} {Phys. Rev. B}\ }\textbf {\bibinfo {volume}
  {65}},\ \bibinfo {pages} {165113} (\bibinfo {year} {2002})}\BibitemShut
  {NoStop}%
\bibitem [{\citenamefont {Kalmeyer}\ and\ \citenamefont
  {Laughlin}(1989)}]{kalmeyer1989theory}%
  \BibitemOpen
  \bibfield  {author} {\bibinfo {author} {\bibfnamefont {V.}~\bibnamefont
  {Kalmeyer}}\ and\ \bibinfo {author} {\bibfnamefont {R.}~\bibnamefont
  {Laughlin}},\ }\href
  {https://journals.aps.org/prb/abstract/10.1103/PhysRevB.39.11879} {\bibfield
  {journal} {\bibinfo  {journal} {Phys. Rev. B}\ }\textbf {\bibinfo {volume}
  {39}},\ \bibinfo {pages} {11879} (\bibinfo {year} {1989})}\BibitemShut
  {NoStop}%
\bibitem [{\citenamefont {Boada}\ \emph {et~al.}(2015)\citenamefont {Boada},
  \citenamefont {Celi}, \citenamefont {Rodr{\'\i}guez-Laguna}, \citenamefont
  {Latorre},\ and\ \citenamefont {Lewenstein}}]{boada2015quantum}%
  \BibitemOpen
  \bibfield  {author} {\bibinfo {author} {\bibfnamefont {O.}~\bibnamefont
  {Boada}}, \bibinfo {author} {\bibfnamefont {A.}~\bibnamefont {Celi}},
  \bibinfo {author} {\bibfnamefont {J.}~\bibnamefont {Rodr{\'\i}guez-Laguna}},
  \bibinfo {author} {\bibfnamefont {J.~I.}\ \bibnamefont {Latorre}}, \ and\
  \bibinfo {author} {\bibfnamefont {M.}~\bibnamefont {Lewenstein}},\ }\href
  {http://iopscience.iop.org/article/10.1088/1367-2630/17/4/045007} {\bibfield
  {journal} {\bibinfo  {journal} {New J. Phys.}\ }\textbf {\bibinfo {volume}
  {17}},\ \bibinfo {pages} {045007} (\bibinfo {year} {2015})}\BibitemShut
  {NoStop}%
\bibitem [{\citenamefont {Grusdt}\ and\ \citenamefont
  {H{\"o}ning}(2014)}]{grusdt2014realization}%
  \BibitemOpen
  \bibfield  {author} {\bibinfo {author} {\bibfnamefont {F.}~\bibnamefont
  {Grusdt}}\ and\ \bibinfo {author} {\bibfnamefont {M.}~\bibnamefont
  {H{\"o}ning}},\ }\href
  {https://journals.aps.org/pra/abstract/10.1103/PhysRevA.90.053623} {\bibfield
   {journal} {\bibinfo  {journal} {Phys. Rev. A}\ }\textbf {\bibinfo {volume}
  {90}},\ \bibinfo {pages} {053623} (\bibinfo {year} {2014})}\BibitemShut
  {NoStop}%
\bibitem [{\citenamefont {Budich}\ \emph {et~al.}(2017)\citenamefont {Budich},
  \citenamefont {Elben}, \citenamefont {{\L}{\k{a}}cki}, \citenamefont
  {Sterdyniak}, \citenamefont {Baranov},\ and\ \citenamefont
  {Zoller}}]{budich2017coupled}%
  \BibitemOpen
  \bibfield  {author} {\bibinfo {author} {\bibfnamefont {J.}~\bibnamefont
  {Budich}}, \bibinfo {author} {\bibfnamefont {A.}~\bibnamefont {Elben}},
  \bibinfo {author} {\bibfnamefont {M.}~\bibnamefont {{\L}{\k{a}}cki}},
  \bibinfo {author} {\bibfnamefont {A.}~\bibnamefont {Sterdyniak}}, \bibinfo
  {author} {\bibfnamefont {M.}~\bibnamefont {Baranov}}, \ and\ \bibinfo
  {author} {\bibfnamefont {P.}~\bibnamefont {Zoller}},\ }\href
  {https://journals.aps.org/pra/abstract/10.1103/PhysRevA.95.043632} {\bibfield
   {journal} {\bibinfo  {journal} {Phys. Rev. A}\ }\textbf {\bibinfo {volume}
  {95}},\ \bibinfo {pages} {043632} (\bibinfo {year} {2017})}\BibitemShut
  {NoStop}%
\bibitem [{\citenamefont {Zupancic}\ \emph {et~al.}(2016)\citenamefont
  {Zupancic}, \citenamefont {Preiss}, \citenamefont {Ma}, \citenamefont
  {Lukin}, \citenamefont {Tai}, \citenamefont {Rispoli}, \citenamefont
  {Islam},\ and\ \citenamefont {Greiner}}]{zupancic2016ultra}%
  \BibitemOpen
  \bibfield  {author} {\bibinfo {author} {\bibfnamefont {P.}~\bibnamefont
  {Zupancic}}, \bibinfo {author} {\bibfnamefont {P.~M.}\ \bibnamefont
  {Preiss}}, \bibinfo {author} {\bibfnamefont {R.}~\bibnamefont {Ma}}, \bibinfo
  {author} {\bibfnamefont {A.}~\bibnamefont {Lukin}}, \bibinfo {author}
  {\bibfnamefont {M.~E.}\ \bibnamefont {Tai}}, \bibinfo {author} {\bibfnamefont
  {M.}~\bibnamefont {Rispoli}}, \bibinfo {author} {\bibfnamefont
  {R.}~\bibnamefont {Islam}}, \ and\ \bibinfo {author} {\bibfnamefont
  {M.}~\bibnamefont {Greiner}},\ }\href
  {https://www.osapublishing.org/oe/abstract.cfm?uri=oe-24-13-13881} {\bibfield
   {journal} {\bibinfo  {journal} {Opt. Express}\ }\textbf {\bibinfo {volume}
  {24}},\ \bibinfo {pages} {13881} (\bibinfo {year} {2016})}\BibitemShut
  {NoStop}%
\bibitem [{\citenamefont {Barredo}\ \emph {et~al.}(2016)\citenamefont
  {Barredo}, \citenamefont {de~L{\'e}s{\'e}leuc}, \citenamefont {Lienhard},
  \citenamefont {Lahaye},\ and\ \citenamefont {Browaeys}}]{barredo2016atom}%
  \BibitemOpen
  \bibfield  {author} {\bibinfo {author} {\bibfnamefont {D.}~\bibnamefont
  {Barredo}}, \bibinfo {author} {\bibfnamefont {S.}~\bibnamefont
  {de~L{\'e}s{\'e}leuc}}, \bibinfo {author} {\bibfnamefont {V.}~\bibnamefont
  {Lienhard}}, \bibinfo {author} {\bibfnamefont {T.}~\bibnamefont {Lahaye}}, \
  and\ \bibinfo {author} {\bibfnamefont {A.}~\bibnamefont {Browaeys}},\ }\href
  {http://science.sciencemag.org/content/354/6315/1021} {\bibfield  {journal}
  {\bibinfo  {journal} {Science}\ }\textbf {\bibinfo {volume} {354}},\ \bibinfo
  {pages} {1021} (\bibinfo {year} {2016})}\BibitemShut {NoStop}%
\bibitem [{\citenamefont {Endres}\ \emph {et~al.}(2016)\citenamefont {Endres},
  \citenamefont {Bernien}, \citenamefont {Keesling}, \citenamefont {Levine},
  \citenamefont {Anschuetz}, \citenamefont {Krajenbrink}, \citenamefont
  {Senko}, \citenamefont {Vuletic}, \citenamefont {Greiner},\ and\
  \citenamefont {Lukin}}]{endres2016atom}%
  \BibitemOpen
  \bibfield  {author} {\bibinfo {author} {\bibfnamefont {M.}~\bibnamefont
  {Endres}}, \bibinfo {author} {\bibfnamefont {H.}~\bibnamefont {Bernien}},
  \bibinfo {author} {\bibfnamefont {A.}~\bibnamefont {Keesling}}, \bibinfo
  {author} {\bibfnamefont {H.}~\bibnamefont {Levine}}, \bibinfo {author}
  {\bibfnamefont {E.~R.}\ \bibnamefont {Anschuetz}}, \bibinfo {author}
  {\bibfnamefont {A.}~\bibnamefont {Krajenbrink}}, \bibinfo {author}
  {\bibfnamefont {C.}~\bibnamefont {Senko}}, \bibinfo {author} {\bibfnamefont
  {V.}~\bibnamefont {Vuletic}}, \bibinfo {author} {\bibfnamefont
  {M.}~\bibnamefont {Greiner}}, \ and\ \bibinfo {author} {\bibfnamefont
  {M.~D.}\ \bibnamefont {Lukin}},\ }\href
  {http://science.sciencemag.org/content/354/6315/1024} {\bibfield  {journal}
  {\bibinfo  {journal} {Science}\ }\textbf {\bibinfo {volume} {354}},\ \bibinfo
  {pages} {1024} (\bibinfo {year} {2016})}\BibitemShut {NoStop}%
\bibitem [{\citenamefont {Schine}\ \emph {et~al.}(2018)\citenamefont {Schine},
  \citenamefont {Chalupnik}, \citenamefont {Can}, \citenamefont {Gromov},\ and\
  \citenamefont {Simon}}]{schine2018measuring}%
  \BibitemOpen
  \bibfield  {author} {\bibinfo {author} {\bibfnamefont {N.}~\bibnamefont
  {Schine}}, \bibinfo {author} {\bibfnamefont {M.}~\bibnamefont {Chalupnik}},
  \bibinfo {author} {\bibfnamefont {T.}~\bibnamefont {Can}}, \bibinfo {author}
  {\bibfnamefont {A.}~\bibnamefont {Gromov}}, \ and\ \bibinfo {author}
  {\bibfnamefont {J.}~\bibnamefont {Simon}},\ }\href@noop {} {\bibfield
  {journal} {\bibinfo  {journal} {arXiv}\ } (\bibinfo {year} {2018})},\ \Eprint
  {http://arxiv.org/abs/1802.04418} {1802.04418} \BibitemShut {NoStop}%
\bibitem [{\citenamefont {Barredo}\ \emph {et~al.}(2017)\citenamefont
  {Barredo}, \citenamefont {Lienhard}, \citenamefont {de~L{\'e}s{\'e}leuc},
  \citenamefont {Lahaye},\ and\ \citenamefont
  {Browaeys}}]{barredo2017synthetic}%
  \BibitemOpen
  \bibfield  {author} {\bibinfo {author} {\bibfnamefont {D.}~\bibnamefont
  {Barredo}}, \bibinfo {author} {\bibfnamefont {V.}~\bibnamefont {Lienhard}},
  \bibinfo {author} {\bibfnamefont {S.}~\bibnamefont {de~L{\'e}s{\'e}leuc}},
  \bibinfo {author} {\bibfnamefont {T.}~\bibnamefont {Lahaye}}, \ and\ \bibinfo
  {author} {\bibfnamefont {A.}~\bibnamefont {Browaeys}},\ }\href@noop {}
  {\bibfield  {journal} {\bibinfo  {journal} {arXiv}\ } (\bibinfo {year}
  {2017})},\ \Eprint {http://arxiv.org/abs/1712.02727} {1712.02727}
  \BibitemShut {NoStop}%
\bibitem [{sup()}]{supplement}%
  \BibitemOpen
  \href@noop {} {\bibinfo  {journal} {See Appendix}\ }\BibitemShut
  {NoStop}%
\bibitem [{\citenamefont {Reed}(2017)}]{mreed}%
  \BibitemOpen
\bibfield  {journal} {  }\bibfield  {author} {\bibinfo {author} {\bibfnamefont
  {M.}~\bibnamefont {Reed}},\ }\href@noop {} {Ph.D. thesis},\ \bibinfo
  {school} {University of Maryland} (\bibinfo {year} {2017})\BibitemShut
  {NoStop}%
\bibitem [{\citenamefont {Grimm}\ \emph {et~al.}(2000)\citenamefont {Grimm},
  \citenamefont {Weidem{\"u}ller},\ and\ \citenamefont
  {Ovchinnikov}}]{grimm2000optical}%
  \BibitemOpen
  \bibfield  {author} {\bibinfo {author} {\bibfnamefont {R.}~\bibnamefont
  {Grimm}}, \bibinfo {author} {\bibfnamefont {M.}~\bibnamefont
  {Weidem{\"u}ller}}, \ and\ \bibinfo {author} {\bibfnamefont {Y.~B.}\
  \bibnamefont {Ovchinnikov}},\ }\href
  {http://www.sciencedirect.com/science/article/pii/S1049250X0860186X}
  {\bibfield  {journal} {\bibinfo  {journal} {Adv. At. Mol. Opt. Phys.}\
  }\textbf {\bibinfo {volume} {42}},\ \bibinfo {pages} {95} (\bibinfo {year}
  {2000})}\BibitemShut {NoStop}%
\bibitem [{\citenamefont {Ryu}\ \emph {et~al.}(2007)\citenamefont {Ryu},
  \citenamefont {Andersen}, \citenamefont {Clade}, \citenamefont {Natarajan},
  \citenamefont {Helmerson},\ and\ \citenamefont
  {Phillips}}]{ryu2007observation}%
  \BibitemOpen
  \bibfield  {author} {\bibinfo {author} {\bibfnamefont {C.}~\bibnamefont
  {Ryu}}, \bibinfo {author} {\bibfnamefont {M.}~\bibnamefont {Andersen}},
  \bibinfo {author} {\bibfnamefont {P.}~\bibnamefont {Clade}}, \bibinfo
  {author} {\bibfnamefont {V.}~\bibnamefont {Natarajan}}, \bibinfo {author}
  {\bibfnamefont {K.}~\bibnamefont {Helmerson}}, \ and\ \bibinfo {author}
  {\bibfnamefont {W.~D.}\ \bibnamefont {Phillips}},\ }\href
  {https://journals.aps.org/prl/abstract/10.1103/PhysRevLett.99.260401}
  {\bibfield  {journal} {\bibinfo  {journal} {Phys. Rev. Lett.}\ }\textbf
  {\bibinfo {volume} {99}},\ \bibinfo {pages} {260401} (\bibinfo {year}
  {2007})}\BibitemShut {NoStop}%
\bibitem [{\citenamefont {Wright}\ \emph {et~al.}(2013)\citenamefont {Wright},
  \citenamefont {Blakestad}, \citenamefont {Lobb}, \citenamefont {Phillips},\
  and\ \citenamefont {Campbell}}]{wright2013threshold}%
  \BibitemOpen
  \bibfield  {author} {\bibinfo {author} {\bibfnamefont {K.}~\bibnamefont
  {Wright}}, \bibinfo {author} {\bibfnamefont {R.}~\bibnamefont {Blakestad}},
  \bibinfo {author} {\bibfnamefont {C.}~\bibnamefont {Lobb}}, \bibinfo {author}
  {\bibfnamefont {W.}~\bibnamefont {Phillips}}, \ and\ \bibinfo {author}
  {\bibfnamefont {G.}~\bibnamefont {Campbell}},\ }\href
  {https://journals.aps.org/pra/abstract/10.1103/PhysRevA.88.063633} {\bibfield
   {journal} {\bibinfo  {journal} {Phys. Rev. A}\ }\textbf {\bibinfo {volume}
  {88}},\ \bibinfo {pages} {063633} (\bibinfo {year} {2013})}\BibitemShut
  {NoStop}%
\bibitem [{\citenamefont {Gross}(1961)}]{gross1961structure}%
  \BibitemOpen
  \bibfield  {author} {\bibinfo {author} {\bibfnamefont {E.~P.}\ \bibnamefont
  {Gross}},\ }\href {https://link.springer.com/article/10.1007/BF02731494}
  {\bibfield  {journal} {\bibinfo  {journal} {Nuovo Cimento}\ }\textbf
  {\bibinfo {volume} {20}},\ \bibinfo {pages} {454} (\bibinfo {year}
  {1961})}\BibitemShut {NoStop}%
\bibitem [{\citenamefont {Pitaevskii}(1961)}]{pitaevskii1961vortex}%
  \BibitemOpen
  \bibfield  {author} {\bibinfo {author} {\bibfnamefont {L.}~\bibnamefont
  {Pitaevskii}},\ }\href
  {http://www.jetp.ac.ru/cgi-bin/e/index/e/13/2/p451?a=list} {\bibfield
  {journal} {\bibinfo  {journal} {Sov. Phys. JETP}\ }\textbf {\bibinfo {volume}
  {13}},\ \bibinfo {pages} {451} (\bibinfo {year} {1961})}\BibitemShut
  {NoStop}%
\bibitem [{\citenamefont {Bao}\ \emph {et~al.}(2003)\citenamefont {Bao},
  \citenamefont {Jaksch},\ and\ \citenamefont {Markowich}}]{bao2003numerical}%
  \BibitemOpen
  \bibfield  {author} {\bibinfo {author} {\bibfnamefont {W.}~\bibnamefont
  {Bao}}, \bibinfo {author} {\bibfnamefont {D.}~\bibnamefont {Jaksch}}, \ and\
  \bibinfo {author} {\bibfnamefont {P.~A.}\ \bibnamefont {Markowich}},\ }\href
  {https://www.sciencedirect.com/science/article/pii/S0021999103001025}
  {\bibfield  {journal} {\bibinfo  {journal} {J. Comput. Phys.}\ }\textbf
  {\bibinfo {volume} {187}},\ \bibinfo {pages} {318} (\bibinfo {year}
  {2003})}\BibitemShut {NoStop}%
\bibitem [{\citenamefont {Hatsugai}\ \emph {et~al.}(1991)\citenamefont
  {Hatsugai}, \citenamefont {Kohmoto},\ and\ \citenamefont
  {Wu}}]{hatsugai1991anyons}%
  \BibitemOpen
  \bibfield  {author} {\bibinfo {author} {\bibfnamefont {Y.}~\bibnamefont
  {Hatsugai}}, \bibinfo {author} {\bibfnamefont {M.}~\bibnamefont {Kohmoto}}, \
  and\ \bibinfo {author} {\bibfnamefont {Y.-S.}\ \bibnamefont {Wu}},\ }\href
  {https://journals.aps.org/prb/abstract/10.1103/PhysRevB.43.10761} {\bibfield
  {journal} {\bibinfo  {journal} {Phys. Rev. B}\ }\textbf {\bibinfo {volume}
  {43}},\ \bibinfo {pages} {10761} (\bibinfo {year} {1991})}\BibitemShut
  {NoStop}%
\bibitem [{\citenamefont {Hafezi}\ \emph {et~al.}(2007)\citenamefont {Hafezi},
  \citenamefont {S{\o}rensen}, \citenamefont {Lukin},\ and\ \citenamefont
  {Demler}}]{hafezi2007characterization}%
  \BibitemOpen
  \bibfield  {author} {\bibinfo {author} {\bibfnamefont {M.}~\bibnamefont
  {Hafezi}}, \bibinfo {author} {\bibfnamefont {A.~S.}\ \bibnamefont
  {S{\o}rensen}}, \bibinfo {author} {\bibfnamefont {M.~D.}\ \bibnamefont
  {Lukin}}, \ and\ \bibinfo {author} {\bibfnamefont {E.}~\bibnamefont
  {Demler}},\ }\href
  {http://iopscience.iop.org/article/10.1209/0295-5075/81/10005} {\bibfield
  {journal} {\bibinfo  {journal} {EPL (Europhys. Lett.)}\ }\textbf {\bibinfo
  {volume} {81}},\ \bibinfo {pages} {10005} (\bibinfo {year}
  {2007})}\BibitemShut {NoStop}%
\bibitem [{\citenamefont {Barkeshli}\ and\ \citenamefont
  {Freedman}(2016)}]{Barkeshli:2016wp}%
  \BibitemOpen
  \bibfield  {author} {\bibinfo {author} {\bibfnamefont {M.}~\bibnamefont
  {Barkeshli}}\ and\ \bibinfo {author} {\bibfnamefont {M.}~\bibnamefont
  {Freedman}},\ }\href
  {https://journals.aps.org/prb/abstract/10.1103/PhysRevB.94.165108} {\bibfield
   {journal} {\bibinfo  {journal} {Phys. Rev. B}\ }\textbf {\bibinfo {volume}
  {94}},\ \bibinfo {pages} {165108} (\bibinfo {year} {2016})}\BibitemShut
  {NoStop}%
\bibitem [{\citenamefont {Zhu}\ \emph {et~al.}(2017)\citenamefont {Zhu},
  \citenamefont {Hafezi},\ and\ \citenamefont {Barkeshli}}]{Zhu:2017tr}%
  \BibitemOpen
  \bibfield  {author} {\bibinfo {author} {\bibfnamefont {G.}~\bibnamefont
  {Zhu}}, \bibinfo {author} {\bibfnamefont {M.}~\bibnamefont {Hafezi}}, \ and\
  \bibinfo {author} {\bibfnamefont {M.}~\bibnamefont {Barkeshli}},\ }\href@noop
  {} {\bibfield  {journal} {\bibinfo  {journal} {arXiv}\ } (\bibinfo {year}
  {2017})},\ \Eprint {http://arxiv.org/abs/1711.05752v1} {1711.05752v1}
  \BibitemShut {NoStop}%
\bibitem [{\citenamefont {Siegman}(1986)}]{siegmanlasers}%
  \BibitemOpen
  \bibfield  {author} {\bibinfo {author} {\bibfnamefont {A.~E.}\ \bibnamefont
  {Siegman}},\ }\href@noop {} {\emph {\bibinfo {title} {Lasers}}}\ (\bibinfo
  {publisher} {University Science Books, Mill Valley, CA},\ \bibinfo {year}
  {1986})\BibitemShut {NoStop}%
\bibitem [{\citenamefont {Robens}\ \emph {et~al.}(2017)\citenamefont {Robens},
  \citenamefont {Brakhane}, \citenamefont {Alt}, \citenamefont {Klei{\ss}ler},
  \citenamefont {Meschede}, \citenamefont {Moon}, \citenamefont {Ramola},\ and\
  \citenamefont {Alberti}}]{robens2017high}%
  \BibitemOpen
  \bibfield  {author} {\bibinfo {author} {\bibfnamefont {C.}~\bibnamefont
  {Robens}}, \bibinfo {author} {\bibfnamefont {S.}~\bibnamefont {Brakhane}},
  \bibinfo {author} {\bibfnamefont {W.}~\bibnamefont {Alt}}, \bibinfo {author}
  {\bibfnamefont {F.}~\bibnamefont {Klei{\ss}ler}}, \bibinfo {author}
  {\bibfnamefont {D.}~\bibnamefont {Meschede}}, \bibinfo {author}
  {\bibfnamefont {G.}~\bibnamefont {Moon}}, \bibinfo {author} {\bibfnamefont
  {G.}~\bibnamefont {Ramola}}, \ and\ \bibinfo {author} {\bibfnamefont
  {A.}~\bibnamefont {Alberti}},\ }\href
  {https://www.osapublishing.org/ol/abstract.cfm?uri=ol-42-6-1043} {\bibfield
  {journal} {\bibinfo  {journal} {Opt. Lett.}\ }\textbf {\bibinfo {volume}
  {42}},\ \bibinfo {pages} {1043} (\bibinfo {year} {2017})}\BibitemShut
  {NoStop}%
\bibitem [{\citenamefont {Gl{\"u}ck}\ \emph {et~al.}(2002)\citenamefont
  {Gl{\"u}ck}, \citenamefont {Kolovsky},\ and\ \citenamefont
  {Korsch}}]{gluck2002wannier}%
  \BibitemOpen
  \bibfield  {author} {\bibinfo {author} {\bibfnamefont {M.}~\bibnamefont
  {Gl{\"u}ck}}, \bibinfo {author} {\bibfnamefont {A.~R.}\ \bibnamefont
  {Kolovsky}}, \ and\ \bibinfo {author} {\bibfnamefont {H.~J.}\ \bibnamefont
  {Korsch}},\ }\href
  {http://www.sciencedirect.com/science/article/pii/S0370157302001424}
  {\bibfield  {journal} {\bibinfo  {journal} {Physics Reports}\ }\textbf
  {\bibinfo {volume} {366}},\ \bibinfo {pages} {103} (\bibinfo {year}
  {2002})}\BibitemShut {NoStop}%
\bibitem [{\citenamefont {Hofstadter}(1976)}]{hofstadter1976energy}%
  \BibitemOpen
  \bibfield  {author} {\bibinfo {author} {\bibfnamefont {D.~R.}\ \bibnamefont
  {Hofstadter}},\ }\href
  {https://journals.aps.org/prb/abstract/10.1103/PhysRevB.14.2239} {\bibfield
  {journal} {\bibinfo  {journal} {Phys. Rev. B}\ }\textbf {\bibinfo {volume}
  {14}},\ \bibinfo {pages} {2239} (\bibinfo {year} {1976})}\BibitemShut
  {NoStop}%
\bibitem [{\citenamefont {Bergholtz}\ and\ \citenamefont
  {Karlhede}(2005)}]{bergholtz2005half}%
  \BibitemOpen
  \bibfield  {author} {\bibinfo {author} {\bibfnamefont {E.~J.}\ \bibnamefont
  {Bergholtz}}\ and\ \bibinfo {author} {\bibfnamefont {A.}~\bibnamefont
  {Karlhede}},\ }\href
  {https://journals.aps.org/prl/abstract/10.1103/PhysRevLett.94.026802}
  {\bibfield  {journal} {\bibinfo  {journal} {Phys. Rev. Lett.}\ }\textbf
  {\bibinfo {volume} {94}},\ \bibinfo {pages} {026802} (\bibinfo {year}
  {2005})}\BibitemShut {NoStop}%
\end{thebibliography}

%

\end{document}